\documentclass[10pt]{article}

\usepackage{fullpage}
\usepackage{CJKutf8} 
\usepackage[utf8]{inputenc}
\usepackage{setspace}
\usepackage{parskip}
\usepackage{titlesec}
\usepackage[section]{placeins}
\usepackage{xcolor}
\usepackage{breakcites}
\usepackage{lineno}
\usepackage{hyphenat}
\usepackage{xcolor}

\usepackage{amsmath}
\usepackage{authblk}
\usepackage{amsfonts}
\usepackage{bm, amssymb}
\usepackage{mathtools}
\usepackage{graphicx}
\usepackage{subcaption}

\DeclareMathOperator*{\arginf}{arg\,inf}
\DeclareMathOperator{\sgn}{sgn}

\newcommand{\real}{\mathbb{R}}
\newcommand{\s}{\ensuremath{\mathbb{S}}}
\newcommand{\ltwo}{\ensuremath{\mathbb{L}^2}}
\newcommand{\inner}[2]{\left\langle#1,#2 \right\rangle}
\newcommand{\dee}{\mathrm{d}}

\PassOptionsToPackage{hyphens}{url}
\usepackage[colorlinks = true,
            linkcolor = blue,
            urlcolor  = blue,
            citecolor = blue,
            anchorcolor = blue]{hyperref}
\usepackage{etoolbox}

\usepackage{natbib}

\renewenvironment{abstract}
  {{\bfseries\noindent{\abstractname}\par\nobreak}\footnotesize}
  {\bigskip}

\titlespacing{\section}{0pt}{*3}{*1}
\titlespacing{\subsection}{0pt}{*2}{*0.5}
\titlespacing{\subsubsection}{0pt}{*1.5}{0pt}

\usepackage{authblk}

\usepackage{graphicx}
\usepackage[space]{grffile}
\usepackage{latexsym}
\usepackage{textcomp}
\usepackage{longtable}
\usepackage{tabulary}
\usepackage{booktabs,array,multirow}
\usepackage{amsfonts,amsmath,amssymb}
\providecommand\citet{\cite}
\providecommand\citep{\cite}

\newif\iflatexml\latexmlfalse

\AtBeginDocument{\DeclareGraphicsExtensions{.pdf,.PDF,.eps,.EPS,.png,.PNG,.tif,.TIF,.jpg,.JPG,.jpeg,.JPEG}}

\usepackage[utf8]{inputenc}
\usepackage[english]{babel}
\usepackage{float}
\usepackage[margin=1.5in]{geometry}

\begin{document}
\begin{CJK}{UTF8}{gbsn}

\title{Elastic Multi-Fidelity Bayesian Model Calibration}

\author[1,2]{J. Derek Tucker}
\author[1]{Gavin Collins}
\author[1,3]{Gabriel Huerta}
\author[1]{Justin L. Brown}

\affil[1]{Statistical Sciences, Sandia National Laboratories}
\affil[2]{Department of Statistics, University of Illinois, Urbana-Champaign}
\affil[3]{Department of Mathematics and Statistics, University of New Mexico}
\vspace{-1em}

  \date{}

\begingroup
\let\center\flushleft
\let\endcenter\endflushleft
\maketitle
\endgroup

\begin{abstract}
Bayesian calibration of functional-output computer models typically relies on dimension reduction techniques, such as functional principal component analysis, which assume that differences among simulator realizations arise only from amplitude variation. When simulator output also exhibits phase variation such as shifts in the timing or location of key features, this assumption is violated. Recent work has addressed this issue through elastic calibration, which aligns functional computer model realizations with observed experimental data prior to dimension reduction. Separately, multi-fidelity methods reduce the cost of calibration by supplementing a small number of expensive high-fidelity simulator runs with a larger ensemble of cheap low-fidelity runs. This is typically done through either a mapping strategy, which corrects low-fidelity predictions toward high-fidelity output, or a fusion strategy, which builds a shared basis across both fidelities. This paper combines these two lines of work, introducing elastic multi-fidelity Bayesian model calibration, which aligns high- and low-fidelity functional output to a common reference before applying multi-fidelity mapping or fusion. On a synthetic two-dimensional calibration problem and a dynamic material properties equation-of-state problem, both elastic multi-fidelity strategies match or improve on the leave-one-out predictive accuracy of a mono-fidelity elastic emulator, with the fusion approach achieving the lowest error. Both strategies also produce tighter calibrated posteriors than the mono-fidelity baseline, with the fusion approach providing the best coverage and parameter estimates closest to the true values.
\end{abstract}

\section{Introduction} \label{sec:intro}
Computer model calibration seeks to infer the physical parameters of a simulator by comparing its output to experimental observations, typically within a Bayesian framework that accounts for both parametric and observational uncertainty. When simulator output is high-dimensional or functional, taking the form of a curve or a field rather than a scalar, the calibration process must first reduce the output to a tractable set of summary quantities before parameters can be inferred. \cite{higdon2008computer} established the standard approach: apply a dimension reduction technique to the functional response, most commonly functional principal component analysis (fPCA), and calibrate against the resulting basis coefficients. This approach implicitly assumes that variation across functional realizations is well-captured by amplitude- or magnitude-only differences. In many physical systems, however, functional responses also vary in the location or timing of their features. For example, a peak may be attained at an earlier or later time, or a transition may occur over a shifted interval, introducing phase variation that standard fPCA conflates with amplitude variation. In such scenarios, the resulting basis is statistically inefficient and can bias the calibrated posterior distribution. \cite{francom:2025} addressed these issues by extending the \cite{higdon2008computer} framework to an elastic setting, aligning functional realizations to a common phase via warping functions prior to dimension reduction, and calibrating over the separated phase and amplitude components.

\begin{figure}[htbp]
    \centering
    \includegraphics[width=\textwidth]{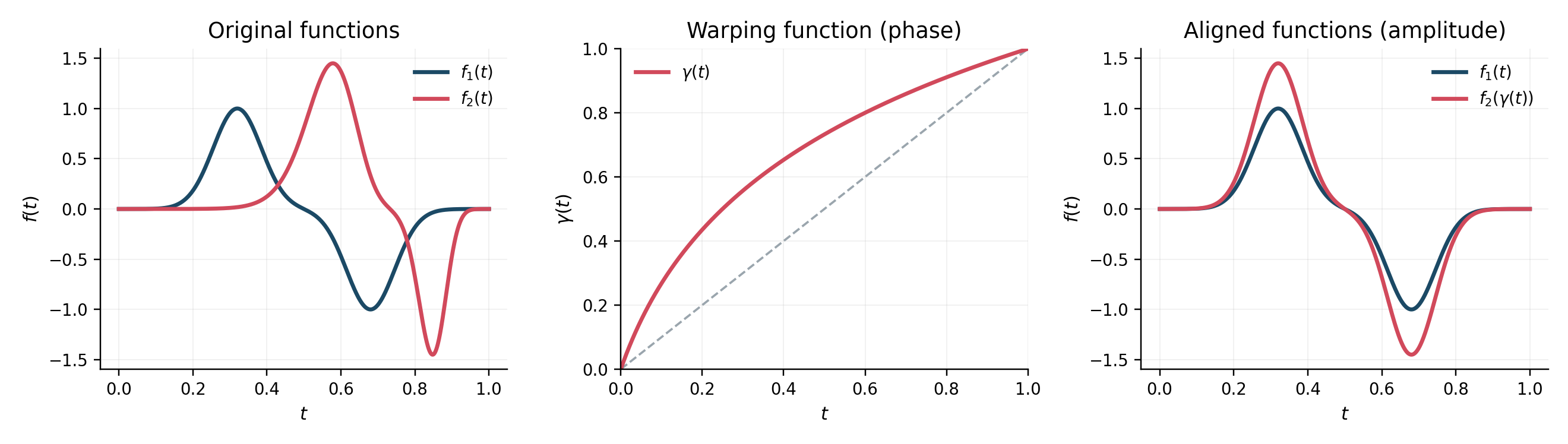}
    \caption{Demonstration of amplitude and phase variability in functional data. Left: original functions. Middle: warping function (``phase variability''). Right: aligned functions (``amplitude variability'').}
    \label{fig:toy_example}
\end{figure}

The motivation for the elastic treatment is a familiar one in functional data analysis, a field with a rich history summarized in \cite{ramsay2005}, \cite{horvath2012}, and \cite{srivastava2016}. The variability in a collection of functional observations can generally be attributed to two sources: \emph{amplitude} variability, which is variation in function values at fixed locations in the function’s domain, and \emph{phase} variability, which is variation in \emph{when} or \emph{where} features occur within the domain. Following \cite{francom:2025}, we use the term \emph{elastic} for methods that handle both sources of variability. Figure~\ref{fig:toy_example} illustrates these concepts on a pair of simulated curves that each contain a peak and a trough. The two curves differ both in the heights of those features (amplitude) and in their locations along the horizontal axis (phase). Applying a transformation of the horizontal axis through a warping function $\gamma$ (middle panel) aligns the curves (right panel); the aligned function $f_1(\gamma(t))$ then encodes the amplitude variability while $\gamma$ itself encodes the phase variability. Standard basis-expansion methods applied to the curves in the left panel must represent horizontal shifts indirectly, using a sequence of vertical adjustments. This is inefficient and, when the resulting residual structure is modeled with a Gaussian likelihood, can also be statistically misleading.

Beyond emulator efficiency, there are three specific reasons to separate the two sources of variability before calibrating \citep{francom:2025}. First, emulation is typically more accurate on aligned data than on misaligned data \citep{francom2022landmark}. Second, model discrepancy can be assigned to the component of the model where it physically belongs: a systematic timing error in the simulator becomes a discrepancy on the phase component, rather than a misplaced amplitude correction that depends on the value of the functional variable. Third, the alignment approach of \cite{srivastava-etal-JASA:2011} supplies a \emph{proper} distance between functions, avoiding the degeneracy, the so-called \emph{pinching effect} \citep{ramsay-li:1998}, that afflicts the $\ltwo$ metric in the native function space. Related deformation-based calibration approaches such as \cite{kleiber2014model} rely on the $\ltwo$ metric and therefore inherit this problem.

A separate obstacle to calibration is computational cost: Bayesian calibration requires many simulator evaluations to characterize the posterior, and high-fidelity (HF) simulators are frequently too expensive to run at the scale required. Multi-fidelity (MF) methods address this by supplementing a small number of HF runs with a much larger set of cheaper, lower-fidelity (LF) runs, using the LF data to reduce the variance or improve the accuracy of quantities estimated primarily from the HF model. A basic taxonomy of multi-fidelity uncertainty quantification (UQ) algorithms begins by distinguishing between sampling-based and surrogate-based approaches, although the two categories are not disjoint. Hybrid methods also exist, such as high-dimensional global sensitivity analysis that uses multi-fidelity sampling to estimate polynomial chaos expansion coefficients. Sampling-based approaches, such as multilevel Monte Carlo, \cite{Giles2008}, multi-fidelity Monte Carlo, \cite{Ng2014, peherstorfer2016optimal}, multilevel-multi-fidelity Monte Carlo, \cite{GeraciCTR, geraci_multi-fidelity_2017}, approximate control variates (ACV), \cite{GORODETSKY2020109257}, and generalized ACV, \cite{Bomarito2022}, share several features. They are robust and dimension-independent with well-defined, exploitable convergence theory; they inherit the (slow) convergence rate of Monte Carlo, since MF variance reduction shifts the convergence curve without changing its rate. Furthermore, extending these approaches to general statistical quantities of interest (e.g., tail probabilities) is non-trivial; and they typically require distinct algorithms for inverse UQ problems, such as multilevel Markov chain Monte Carlo, \cite{siam_ml_mcmc}.

Multi-fidelity surrogate-based UQ approaches grew out of multi-fidelity surrogate-based optimization as in \cite{Eld04, Eld06b, Rob06a, Rob06b}, with early work focused on discrepancy modeling using polynomial chaos and stochastic collocation \cite{NgEldred2012, Eldred2017} and, more recently, with functional tensor train approaches,\cite{gorodetsky2019continuous}. These formulations have increasingly incorporated data-driven surrogates, projection-based reduced-order models, Gaussian processes, and deep neural networks. Relative to sampling-based methods, surrogate-based MF approaches converge rapidly when the input dimension is low to moderate and the quantities of interest are smooth functions of the uncertain inputs. They can exploit special structure such as anisotropy, sparsity, low rank, or reduced-dimensional manifolds. Additionally, these approaches generalize readily across both statistical targets and forward/inverse UQ problems. When the surrogate model must represent functional output specifically, two main strategies emerge. The first is mapping, which involves training a surrogate on the low-fidelity (LF) data and then using a second surrogate to correct or map LF predictions to the high-fidelity (HF) response. The second strategy is fusion, which builds a single enhanced basis spanning both the HF and LF samples and fits surrogates to the resulting coefficients, \cite{benamara2017multi-fidelity-c96}. The recent survey by \cite{BRUNEL2025117577} provides a unification of this body of work.

Despite all these developments, existing multi-fidelity surrogate methods for functional-output simulators are based on the same amplitude-only assumption as in \cite{higdon2008computer}. These methods build their basis on a reduced dimension and fit their emulators directly on the raw functional data, without accounting for phase variation between HF and LF simulator output. This variation that can differ systematically between fidelity levels, since simplified or coarsened models often shift the timing of physical features even when they approximate their magnitude reasonably well. No existing multi-fidelity framework combines the elastic, phase-amplitude separated representation of \cite{francom:2025} with multi-fidelity fusion or mapping.

The key goal of this paper is to close that gap. We introduce an elastic multi-fidelity Bayesian model calibration framework, which aligns the HF and LF functional simulator outputs to a common reference before fitting multi-fidelity surrogates, and extends both the mapping and fusion strategies to operate on the resulting phase (warping function) and amplitude (aligned function) components. We evaluate the proposed approaches on a synthetic two-dimensional calibration problem with a large low-fidelity ensemble ($N = 150$) and a small high-fidelity ensemble ($n = 20$), comparing them to a mono-fidelity elastic calibration baseline that uses only the HF data. Using leave-one-out cross-validation, we show that both elastic MF strategies match or improve on the mono-fidelity emulator's predictive accuracy, with the fusion approach achieving the lowest error on average. We further show that both MF approaches tighten the calibrated posterior relative to the mono-fidelity baseline, with the fusion approach yielding the best coverage and the posterior closest to the true parameter values. This demonstrates that incorporating properly aligned less costly low-fidelity functional data, can meaningfully improve calibration when only a limited high-fidelity budget is available.

We then apply the methodology to a dynamic materials problem in which both obstacles, misalignment and simulator cost are present. We calibrate an equation-of-state (EoS) of tin from experiments conducted on Sandia National Laboratories' Z-machine \citep{brown2023, savage2007}, a pulsed power driver that delivers very large electric currents over short time scales and thereby, drives a time-dependent stress wave through a material sample. The measured response is a velocimetry curve, a functional output whose shock features arrive at times that depend on the material parameters. Therefore, the calibration parameters induce phase variability as well as amplitude variability on the resulting curves. In relation to these experiments, there is a simulator which is a hydrodynamics code whose cost is governed largely by mesh resolution. The code is considered numerically converged at roughly a one-micron mesh, while a ten-micron mesh runs faster by a factor of 25. This yields a natural fidelity hierarchy, with a limited set of converged HF runs and a much larger set of coarse LF runs. This is exactly the regime for which the proposed method is designed. We show that using the LF ensemble improves coverage of the shock region defined by the velocimetry curves and tightens the posterior distributions of the EoS parameters relative to calibrating on the HF runs alone. 

The remainder of the paper is organized as follows. Section~\ref{sec:cal} reviews Bayesian model calibration for univariate and functional response and then develops the elastic approach, covering elastic functional data analysis, the construction of proper amplitude and phase distances, and the resulting calibration model. Section~\ref{sec:mf} describes the two multi-fidelity surrogate strategies, mapping and fusion, in the elastic setting. Section~\ref{sec:results} presents results on the synthetic two-dimensional problem and on the tin Z-machine calibration. Section~\ref{sec:discussion} offers conclusions and directions for future work.

\section{Functional Bayesian Model Calibration}
\label{sec:cal}

Bayesian model calibration seeks to infer the uncertain inputs of a computer model by comparing its output to experimental measurements, while accounting for the several sources of uncertainty that enter that comparison. Following \cite{kennedy2001bayesian}, these sources include parameter uncertainty, measurement uncertainty, emulator or surrogate uncertainty, and model form error (also called model discrepancy or model bias). We first review the standard formulation for univariate response and its extension to functional response. Since the multi-fidelity (MF) constructions of Section~\ref{sec:mf} operate within the functional framework, we then briefly describe the elastic approach of \cite{francom:2025} that we rely upon for the MF construction.

\subsection{Univariate Response}
\label{sec:univariate}

Let $y(\bm x, \bm \theta)$ generally denote the computer model output, where $\bm x$ denotes known conditions that are often fixed settings of an experiment, and $\bm \theta$ denotes the uncertain calibration parameters. Let $z(\bm x)$ denote an experimental observation at $\bm x$, and let $n$ be the number of observations, so that the observed data are $z(\bm x_1), \dots, z(\bm x_n)$. The calibration model of \cite{kennedy2001bayesian} is
\begin{align}
z(\bm x_i) &= y(\bm x_i, \bm\theta) + \delta(\bm x_i) + \epsilon(\bm x_i), \quad \epsilon(\bm x_i) \sim \mathcal{N}(0, \sigma_\epsilon^2), \label{eq:koh1}
\end{align}
where $\delta$ denotes latent error in model form or model discrepancy, and $\epsilon (\bm x)$ denotes measurement error in $z(\bm x)$. The errors are assumed Gaussian, independent and identically distributed, though other error structures may be considered. The unknowns in the calibration model are $\bm\theta$, the measurement error variance $\sigma_\epsilon^2$, and the discrepancy function $\delta(\bm x)$.  

Identifiability requires care within this modeling framework. When $n$ is small, the prior for $\sigma_\epsilon^2$ is influential for the calibration results. Furthermore, there is a natural trade-off between the calibration parameters and the model discrepancy, so that at least one of the two must be well-constrained \textit{a priori} and in particular, the calibration parameters $\bm \theta$. Moreover, a common choice for the discrepancy function $\delta(\bm x)$ is a Gaussian process prior that favors small, smooth functions, or a prior that enforces constraints such as positivity or monotonicity \citep{higdon2004combining, bryn2014}. Typically the covariance function of this Gaussian process is assumed isotropic, i.e., specific covariance values depend on distances between the inputs $\bm x$.

In most realistic settings, the model $y(\bm x, \bm \theta)$ is computationally expensive to evaluate, which makes direct evaluation of the likelihood function inside a Markov chain Monte Carlo (MCMC) sampler impractical. In these cases, a surrogate model, or emulator, is trained on a relatively small design of model evaluations $\{y(\bm x_j, \bm \theta_j)\}_{j=1}^{N_{sim}}$ and used in place of $y(\bm x,\bm \theta)$ \citep{sacks1989SS}. A full Bayesian treatment of the emulator/surrogate combined the the calibration model learns all unknowns, including surrogate parameters, conditional on both the $n$ observations and the $N_{sim}$ model runs, \citep{higdon2004combining}. On the other hand, \cite{kennedy2001bayesian} recommend to fix some emulator and discrepancy parameters before estimating the calibration parameters with MCMC to deal with computational and identifiability issues. Furthermore, \cite{liu2009modularization} describe cases in which modularizing the model, i.e., fitting the emulator first and then treating it as fixed within the calibration, yields more robust inferences in $\bm \theta$. We adopt a modularization strategy for the multifidelity framework proposed in this paper.

\subsection{Functional Response}
\label{sec:pointwise_cal}

Computer model outputs are frequently functional, measured over space and/or time. The majority of early calibration work on functional outputs considered extracting scalar features from the curves, such as peak values or critical points \citep{walters2018bayesian}, but feature extraction could be tedious, error prone, and problem-specific. Therefore, direct approaches for multivariate or functional response have been developed \citep{higdon2008computer, bayarri2007framework, francom2019inferring}. Functional responses can vastly increase the training data size to implement surrogate modeling withing a calibration framework, though Gaussian processes with Kronecker covariance structures can produce scalable approaches \citep{williams2006combining}. Alternatively, \cite{gu2016parallel} proposed to fit a Gaussian process for each functional output while sharing parameters across models.

The most common approach for calibration with functional outputs projects the functional response onto a set of basis functions and performs calibration in the reduced-dimension coefficient space \citep{higdon2008computer, bayarri2007framework, francom2019inferring}. Let $t$ denote the functional variable (e.g., time), let $z(t, \bm x_i)$ denote the experimental measurement from experiment $i$, and let $y(t, \bm x_i, \bm \theta)$ denote a simulation of experiment $i$ at input parameters $\bm u$. The calibration model becomes
\begin{align}
z(t, \bm x_i) &= y(t, \bm x_i, \bm\theta) + \delta(t, \bm x_i) + \epsilon(t, \bm x_i), \quad \epsilon(t, \bm x_i) \sim \mathcal{N}(0, \sigma_\epsilon^2),
\end{align}
and inference under this model proceeds by discretizing $t$ onto a grid $t_1, \dots, t_{N_T}$, which produces $N_T$-dimensional vectors $\bm z(\bm x_i)$, $\bm y(\bm x_i, \bm\theta)$, and $\bm\delta(\bm x_i)$ and reduces the model to the multivariate form
\begin{align}
\bm z(\bm x_i) &= \bm y(\bm x_i, \bm\theta) + \bm\delta(\bm x_i) + \bm\epsilon(\bm x_i), \quad \bm\epsilon(\bm x_i) \sim \mathcal{N}(\bm 0, \sigma_\epsilon^2 \bm I).
\end{align}
As in the univariate case, an emulator is required when $y(\bm x_i, \bm \theta)$ is computationally expensive to evaluate. \cite{higdon2008computer} proposed to project the model runs onto functional principal components (fPCA) 
onto a separate flexible basis, carrying out inference in the resulting low-dimensional space. Also \cite{francom2019inferring} proposed to project the model runs onto functional principal components, allowing the discrepancy to be fully dimensional, and fit the emulator in a modular fashion. Alternatively \cite{bayarri2007framework} proposed to use a wavelet basis to reduce dimensionality. 

All of these approaches do not consider the possibility that the computer model output corresponds to misaligned functional data. Therefore, these methods can have challenges to deal with both emulation and model discrepancy since they only account for amplitude variation of the functional outputs. Within the calibration process, these methods compute dissimilarity values between functions which are not proper distances, given that they ignore phase variability.

\subsection{Elastic Approach}
\label{sec:elastic}

The elastic approach of \cite{francom:2025} resolves issues for calibration with misaligned outputs by decomposing each misaligned functional response into an aligned function and a warping function, and then reformulating the calibration on the two resulting datasets rather than on the original responses. 

Subsections \ref{sec:align}, \ref{sec:alignment} and \ref{sec:phase} review the elements of elastic functional data analysis (EFDA) that produce the decomposition that captures amplitude and phase variability and defines an appropriate distance for misaligned curves. Subsection \ref{sec:ebcalibration} reviews the calibration model resulting from the elastic approach. 

\subsubsection{Types of Variability and Model Discrepancy}
\label{sec:align}

When model outputs are functional, two types of variability must be considered. Amplitude variability is variability in the output at a fixed value of the functional variable, that is, $y$-axis variation for one-dimensional functions. Phase variability is variability in the functional variable itself, that is, $x$-axis variation. Input variables can induce both, producing outputs with fundamentally different shapes across the range of plausible inputs.

Model discrepancy compounds the issue. Even at the correct value of the model inputs, an imperfect model will not match the observed data. In the case of misaligned functional data, this mismatch may be in either phase, amplitude, or both. Representing discrepancy in both spaces is therefore necessary to faithfully describe discrepancy-induced shape distortions. When discrepancy is not driven solely by amplitude variability, the pointwise calibration of Section~\ref{sec:pointwise_cal} produces biased estimates of the calibration parameters.

A further problem with relying on the $\ltwo$ metric in the original function space is the \emph{pinching problem} \citep{ramsay-li:1998}: if two functions $f_1$ and $f_2$ are such that the range of $f_1$ lies entirely above the range of $f_2$, the $\ltwo$ metric becomes degenerate and pinches the warped function. To address this, \cite{srivastava-etal-JASA:2011} introduced a mapping for functional data called the \emph{square-root velocity function} (SRVF), which improves functional alignment and supplies the mathematical equalities underlying this topic. The metric used in the alignment is a proper distance and avoids the pinching effect without recourse to a penalty.

\subsubsection{Functional Alignment}
\label{sec:alignment}

To describe the metrics used in calibration, it suffices to consider the comparison of two functions of $t$, namely $z(t, \bm x_i)$ and $y(t, \bm x_i, \bm\theta)$. Varying $\bm\theta$ changes the shape of $y(t, \bm x_i, \bm\theta)$, and we seek the value of $\bm\theta$ at which the two curves are \emph{optimally matched}, where optimality accounts for distance in both amplitude and phase. For notational convenience we suppress $\bm x_i$ and $\bm\theta$ in this subsection and use $z(t)$ and $y(t)$ to denote the functions.

The premise of EFDA \citep{srivastava2016} is to construct a proper distance between the computational prediction $y(t)$ and the experimental data $z(t)$. A continuous mapping $\gamma_{y \rightarrow z}(t): [0,1] \rightarrow [0,1]$ is constructed such that $\gamma$ is a diffeomorphism. The function $\gamma_{y \rightarrow z}$ is called a warping function, since it measures phase distortion in $y$ such that $y \circ \gamma_{y \rightarrow z}(t) = y(\gamma_{y \rightarrow z}(t))$ aligns with $z(t)$. \cite{srivastava-etal-JASA:2011} and \cite{tucker-wu-srivastava:2013} show that applying a particular transformation to $z$ and $y$ yields simple expressions for the amplitude and phase distance between them.

The functions are transformed to their SRVFs. For a function $f(t)$, the SRVF is defined as
\begin{align}
q(t) &= \sgn(\dot f(t)) \sqrt{|\dot f(t)|},
\end{align}
where $\dot f$ denotes the time derivative of $f$. The SRVF is a bijective mapping up to translation: $f(t)$ can be recovered uniquely from $q(t)$ together with a single point on the curve.

Let $q_z(t)$ and $q_y(t)$ denote the SRVFs of $z(t)$ and $y(t)$. The warping function aligning $y$ to $z$ is obtained by solving
\begin{align}
\label{eq:optim}
\gamma_{y \rightarrow z} = \arginf_{\gamma \in \Gamma} \left\| q_z - (q_y \circ \gamma)\sqrt{\dot\gamma} \right\|,
\end{align}
which can be computed by dynamic programming \citep{tucker-wu-srivastava:2013} or through a Bayesian formulation \citep{cheng2016bayesian, lu2017bayesian}. An advantage of this construction is that the analyst does not to specify landmarks; estimation of $\gamma$ exploits the group structure of $\Gamma$. We refer to the use of \eqref{eq:optim} as the \emph{warping decomposition}, since it decomposes a misaligned function into an aligned function and a warping function.

Given $\gamma_{y \rightarrow z}$ and the aligned SRVF $(q_y \circ \gamma_{y \rightarrow z})\sqrt{\dot\gamma_{y \rightarrow z}}$, amplitude variability is measured by
\begin{align}
d_a(q_z, q_y) &= \left\| q_z - (q_y \circ \gamma_{y \rightarrow z})\sqrt{\dot\gamma_{y \rightarrow z}} \right\|^2 .
\end{align}
\cite{srivastava-etal-JASA:2011} show that this $\ltwo$ distance on the transformed and aligned SRVFs is a proper distance metric for amplitude.

\subsubsection{Representing and Measuring Phase Variability}
\label{sec:phase}

Defining a measure of phase variability is more intricate than for amplitude, because the space of warping functions $\Gamma$ is an infinite-dimensional nonlinear manifold and cannot be treated as a Hilbert space. Some transformation of $\Gamma$ into a linear space is therefore required before warping functions can be emulated, differenced, projected onto a basis, or assigned a Gaussian likelihood function, all of which are demanded by the calibration and multi-fidelity constructions of this paper.

A standard element in elastic FDA is to represent $\gamma \in \Gamma$ by the square root of its derivative, $\psi = \sqrt{\dot\gamma}$, which is the SRVF of $\gamma$ and takes this simplified form because $\dot\gamma > 0$. Since $\int_0^1 \psi(t)^2 \dee t = \int_0^1 \dot\gamma(t)\, \dee t = \gamma(1) - \gamma(0) = 1$, the set of all such $\psi$ is the positive orthant of the Hilbert sphere $\s_\infty^+$, and the phase distance is the arc length between representations on that sphere. Because the sphere is still a nonlinear manifold, modeling proceeds in the tangent space at the identity, and the resulting tangent-space representations are the \emph{shooting vectors} used by \cite{francom:2025}.

This construction has a drawback that becomes acute in the multi-fidelity setting. The tangent space is a local linearization: the warping functions correspond only to a bounded region of the positive orthant, and linear operations performed in the tangent space can leave that region, in which case the inverse-exponential map returns objects that are not valid warping functions at all, being non-monotone \citep{happ2019}. The elastic multi-fidelity constructions of Section~\ref{sec:mf} are built entirely from linear operations on the phase representation, differences between fidelity levels, projections onto an enhanced basis, and residuals from those projections, so any representation whose admissible set is not closed under such operations has a poor foundation. We therefore adopt the linear representation of \cite{ma2024stochastic}, which is built on the log-derivative, or the centered log-ratio (CLR) transform.

Following \cite{ma2024stochastic}, we restrict attention to the set of warping functions with bounded derivative,
\begin{align}
\Gamma_1 = \left\{\gamma: [0,1] \rightarrow [0,1] ~\middle|~ \gamma(0) = 0,~ \gamma(1) = 1,~ 0 < m_\gamma < \dot\gamma(t) < M_\gamma < \infty \right\},
\end{align}
which excludes only degenerate warpings and is no restriction in practice, since the warping functions returned by the regularized decomposition of Section~\ref{sec:alignment} have bounded derivative by construction. The set $\Gamma_1$ is not a vector space under the conventional $\ltwo$ metric, but \cite{ma2024stochastic} equip it with perturbation and power operations,
\begin{align}
[\gamma_1 \oplus_\Gamma \gamma_2](t) = \frac{\int_0^t \dot\gamma_1(s)\dot\gamma_2(s)\, \dee s}{\int_0^1 \dot\gamma_1(\tau)\dot\gamma_2(\tau)\, \dee \tau}, \qquad
[\alpha \odot_\Gamma \gamma](t) = \frac{\int_0^t \dot\gamma^\alpha(s)\, \dee s}{\int_0^1 \dot\gamma^\alpha(\tau)\, \dee \tau},
\end{align}
for $\gamma, \gamma_1, \gamma_2 \in \Gamma_1$ and $\alpha \in \real$, together with the inner product
\begin{align}
\label{eq:clr_inner}
\inner{\gamma_1}{\gamma_2}_\Gamma = \int_0^1 \log\left(\dot\gamma_1(t)\right)\log\left(\dot\gamma_2(t)\right) \dee t - \int_0^1 \log\left(\dot\gamma_1(s)\right) \dee s \int_0^1 \log\left(\dot\gamma_2(t)\right) \dee t,
\end{align}
under which $\Gamma_1$ becomes an inner-product space with an associated norm and metric.

The representation itself is the CLR transform of the derivative. Define $\mathcal{C}: \Gamma_1 \rightarrow H(0,1)$ by
\begin{align}
\label{eq:clr}
h(t) = \mathcal{C}(\gamma)(t) = \log\left(\dot\gamma(t)\right) - \int_0^1 \log\left(\dot\gamma(s)\right) \dee s,
\end{align}
where the image space is
\begin{align}
H(0,1) = \left\{h \in \ltwo([0,1]) ~\middle|~ \int_0^1 h(t)\, \dee t = 0,~ -\infty < m_h < h(t) < M_h < \infty\right\}.
\end{align}
\cite{ma2024stochastic} show that $\Gamma_1$ and $H(0,1)$ are isometrically isomorphic under the operations above, and that the inverse mapping is
\begin{align}
\label{eq:clr_inv}
\gamma(t) = \mathcal{C}^{-1}(h)(t) = \frac{\int_0^t \exp(h(s))\, \dee s}{\int_0^1 \exp(h(\tau))\, \dee \tau}.
\end{align}
Removing the point-wise bounds gives the smallest Hilbert space containing $H(0,1)$,
\begin{align}
E(0,1) = \left\{h \in \ltwo([0,1]) ~\middle|~ \int_0^1 h(t)\, \dee t = 0 \right\},
\end{align}
which is a closed linear subspace of $\ltwo([0,1])$ defined by the single linear constraint that $h$ integrate to zero. A complete orthonormal system for $E(0,1)$ is obtained by deleting the constant element from the Fourier basis,
\begin{align}
B = \left\{\phi_{2j-1}(t) = \sqrt{2}\sin(2j\pi t),~ \phi_{2j}(t) = \sqrt{2}\cos(2j\pi t),~ j \ge 1,~ t \in [0,1]\right\},
\end{align}
though any mean-zero basis may be used, including one estimated from the observed $h$ by fPCA.

Because $\mathcal{C}$ is an isometry, the phase distance is simply the $\ltwo$ distance between representations,
\begin{align}
\label{eq:phase_dist}
d_p(\gamma_1, \gamma_2) = \left\| \mathcal{C}(\gamma_1) - \mathcal{C}(\gamma_2) \right\| = \left\| h_1 - h_2 \right\|,
\end{align}
which is a proper distance and is exactly the metric induced by \eqref{eq:clr_inner}. Three properties of this representation matter for what follows.

First, and most importantly, $E(0,1)$ is a genuine linear subspace rather than a tangent-space approximation to a curved one. Sums, differences, scalar multiples, and basis projections of elements of $E(0,1)$ remain in $E(0,1)$, and \eqref{eq:clr_inv} maps any such element back to a strictly increasing warping function, since $\exp(h) > 0$ everywhere. There is no admissible region to leave and no possibility of producing a non-monotone result. The multi-fidelity operations of Section~\ref{sec:mf} are therefore well-defined on the phase component without qualification. The emulator predictions and discrepancy corrections in $h$-space always correspond to valid warping functions.

Second, the identity warping $\gamma_{id}(t) = t$ has $\dot\gamma_{id} \equiv 1$ and therefore maps to $h \equiv 0$, the origin of $E(0,1)$, with norm zero \citep{ma2024stochastic}. Because the simulations are aligned to the experimental data $z$, the warping function for the experiment is the identity $\gamma_{z \rightarrow z}$ and its representation is exactly zero. Deviations of $\bm h$ from zero therefore measure departure in phase from the experimental data, and the norm $\|\bm h\|$ grows as the warping departs further from the identity. This is the property exploited in Section~\ref{sec:results}, where simulations that align well with the experiment produce warping functions near the identity and log-derivative representations near zero. It is the same interpretive convenience that the shooting vector provides, obtained here without the tangent-space construction.

Third, the map \eqref{eq:clr_inv} is invariant to the addition of a constant, since $\exp(h + c) = e^c \exp(h)$ leaves the ratio unchanged; the zero-mean constraint is precisely what fixes a unique representative in each equivalence class. A practical consequence, which mirrors the corresponding remark for shooting vectors in \cite{francom:2025}, is that a constant offset in $h$-space has no effect on the recovered warping function, so a constant time shift cannot be encoded by adding a constant to $\bm h$. Discrepancy in phase must instead be expressed through basis elements with non-constant shape, for which the mean-zero Fourier system $B$ or a piecewise linear mean-zero basis are natural choices.

\subsubsection{Elastic Bayesian Model Calibration}
\label{sec:ebcalibration}

Based on proper amplitude and phase distances, the calibration proceeds by decomposing both the experimental and simulated curves. The observation from experiment $i$ is written as
\begin{align}
z(t, \bm x_i) &= \tilde z(t, \bm x_i) \circ_t \gamma_{z \rightarrow z}(t, \bm x_i),
\end{align}
where $\circ_t$ emphasizes that composition acts only in $t$, so that $f(a,b) \circ_a g(a,c) = f(g(a,c), b)$. This is the identity warping, but stating it explicitly permits the generalizations discussed below. Each simulation $j$ is decomposed similarly using \eqref{eq:optim},
\begin{align}
y(t, \bm x_j, \bm \theta_j) &= \tilde y(t, \bm x_j, \bm \theta_j) \circ_t \gamma_{y \rightarrow z}(t, \bm x_j, \bm \theta_j).
\end{align}
To model with proper distances, the warping functions are transformed by \eqref{eq:clr} into the linear space $E(0,1)$,
\begin{align}
\bm h_{z \rightarrow z}(\bm x_i) &= \mathcal{C}\left(\gamma_{z \rightarrow z}(\bm x_i)\right), \\
\bm h_{y \rightarrow z}(\bm x_j, \bm \theta_j) &= \mathcal{C}\left(\gamma_{y \rightarrow z}(\bm x_j, \bm \theta_j)\right).
\end{align}
Since $\gamma_{z \rightarrow z}$ is the identity, $\bm h_{z \rightarrow z}(\bm x_i) = \bm 0$. The aligned model runs $\tilde y$ and the associated log-derivative representations $\bm h_{y \rightarrow z}$ are the two datasets on which emulators are built, and they are precisely the quantities to which the multi-fidelity strategies of Section~\ref{sec:mf} are applied.

The calibration model using the aligned data and the log-derivative representations has likelihood components
\begin{align}
\tilde z(t, \bm x_i) &= \tilde y(t, \bm x_i, \bm\theta) + \delta_{\tilde y}(t, \bm x_i) + \epsilon_{\tilde z}(t, \bm x_i), \quad \epsilon_{\tilde z}(t, \bm x_i) \sim \mathcal{N}(0, \sigma_{\tilde z}^2), \label{eqn:ye} \\
\bm h_{z \rightarrow z}(\bm x_i) &= \bm h_{y \rightarrow z}(\bm x_i, \bm\theta) + \bm\delta_h(\bm x_i) + \bm\epsilon_h(\bm x_i), \quad \bm\epsilon_h(\bm x_i) \sim \mathcal{N}(\bm 0, \sigma_h^2 \bm I). \label{eqn:ve}
\end{align}
Discretizing $t$ allows \eqref{eqn:ye} to be written in vector form, $\tilde{\bm z}(\bm x_i) = \tilde{\bm y}(\bm x_i, \bm\theta) + \bm\delta_{\tilde y}(\bm x_i) + \bm\epsilon_{\tilde z}(\bm x_i)$, and the joint likelihood is then

\begin{equation} \label{eqn:likelihood}
\begin{aligned}
f\left(\tilde{\bm z}(\bm x_1), \dots, \tilde{\bm z}(\bm x_n), \bm h_{z \rightarrow z}(\bm x_1), \dots, \bm h_{z \rightarrow z}(\bm x_n) ~\middle|~ \bm\theta, \sigma_{\tilde z}^2, \sigma_h^2, \bm\beta_{\tilde y}, \bm\beta_h\right) = \\
\prod_{i=1}^n \left[ \mathcal{N}\left(\tilde{\bm z}(\bm x_i) ~\middle|~ \tilde{\bm y}(\bm x_i, \bm\theta) + \bm\delta_{\tilde y}(\bm x_i),~ \sigma_{\tilde z}^2 \bm I\right) \mathcal{N}\left(\bm h_{z \rightarrow z}(\bm x_i) ~\middle|~ \bm h_{y \rightarrow z}(\bm x_i, \bm\theta) + \bm\delta_h(\bm x_i),~ \sigma_h^2 \bm I\right) \right],
\end{aligned}
\end{equation}

where $\bm\beta_{\tilde y}$ and $\bm\beta_h$ parameterize the discrepancy. Because $d_p$ in \eqref{eq:phase_dist} is an $\ltwo$ distance in $E(0,1)$, the Gaussian kernel in the second factor is a proper phase distance without further transformation. With priors for $\bm\theta$, $\sigma_{\tilde z}^2$, $\sigma_h^2$, $\bm\beta_{\tilde y}$, and $\bm\beta_h$, the posterior is proportional to the likelihood times the priors and can be sampled by MCMC.

When emulation is unnecessary, this formulation still improves on point-wise calibration since the discrepancy model can be specified more sensibly and the likelihood function uses a proper distance. In this case, calibration does, however, require a warping decomposition at every likelihood evaluation. Although decomposing a single curve is fast, the cost accumulates over MCMC runs, which makes emulation attractive even for inexpensive simulators. When emulation is used, separate or joint emulators are trained with inputs $\{\bm x_j, \bm \theta_j\}_{j=1}^{N_{sim}}$ and outputs $\{\tilde{\bm y}(\bm x_j, \bm \theta_j), \bm h_{y \rightarrow z}(\bm x_j, \bm \theta_j)\}_{j=1}^{N_{sim}}$; a full Bayesian treatment forms a joint likelihood over the observations and the model runs, while a modular treatment fits the emulators first and treats them as fixed within the calibration. Since the resulting likelihoods are similar to Equation \eqref{eqn:likelihood}, we omit them for brevity. The corresponding expressions for Gaussian process emulators appear in \cite{higdon2004combining} and \cite{higdon2008computer}. All the results presented in this paper rely on the modular formulation. For additional details and practical considerations, please see details in \cite{francom:2025}.

\section{Multi-Fidelity Approaches}
\label{sec:mf}

Multi-fidelity surrogate modeling techniques can be broadly categorized into two main approaches: mapping and fusion. The mapping approach begins by training a surrogate model using low-fidelity data, followed by additional models that learn to transform or correct these low-fidelity predictions to approximate high-fidelity results. In contrast, the fusion approach integrates both high- and low-fidelity data simultaneously by constructing a shared representation or basis that captures information from both sources.

At any input setting $\bm{x}$, we decompose the simulator output $y(\bm{x})$ into an emulator prediction $\hat{y}(\bm{x})$ and a remaining error term $\epsilon(\bm{x})$:
\begin{equation}\label{eq:surr}
y(\bm{x}) = \hat{y}(\bm{x}) + \epsilon(\bm{x}).
\end{equation}

A standard assumption is that this error term follows a zero-mean Gaussian distribution,
\begin{equation}
\epsilon(\bm{x}) \sim \mathcal{N}\!\left(0, v(\bm{x})\right),
\end{equation}
with the input-dependent variance $v(\bm{x})$ quantifying the residual uncertainty.

We assume that we are able to run the simulator in a high-fidelity (HF) setting for a small number of samples, $n$, at the input points $\bm{x}_{HF}$. Additionally, we assume we can run the simulator at a lower fidelity (LF) setting to more quickly generate a large number of samples, $N$, at input points $\bm{x}_{LF}$. For each of these cases, we can construct a surrogate approximation $y_{HF}(\bm{x}_{HF})$ and $y_{LF}(\bm{x}_{LF})$, respectively.

In the elastic setting, neither the mapping nor fusion strategies are applied directly to the raw simulator output. Instead, the HF and LF ensembles are first passed through the warping decomposition of Section~\ref{sec:ebcalibration}, producing aligned functions $\tilde y$ and log-derivative representations $\bm h$ for each fidelity level. The mapping or fusion construction described below is then carried out twice: once on the aligned functions and once on the $\bm h$ functions. Two points are worth emphasizing. First, both fidelity levels must be aligned to a \emph{common} reference, taken here to be the experimental curve $z$, so that the aligned functions occupy a common amplitude space and the log-derivative representations a common phase space. Aligning each fidelity to its own reference would make the HF and LF representations incommensurable and the difference and residual quantities defined below meaningless. Second, because $E(0,1)$ is a closed linear subspace of $\ltwo$, the linear-algebraic operations underlying both strategies---differencing, projection onto a basis, and residual formation---are well defined in this space and produce elements that remain in $E(0,1)$. This approach enables the phase component to be handled in the same manner as the amplitude component, motivating the use of the representation from Section~\ref{sec:phase} instead of a tangent-space representation, whose admissible set is not closed under these operations. Prediction of a HF curve then requires recovering the warping function from the predicted $\bm h$ through \eqref{eq:clr_inv} and composing it with the predicted aligned function, returning the prediction to the original data space.

\subsection{Mapping}
For the mapping approach, we first build a surrogate model on $y_{LF}$ utilizing fPCA as the chosen dimension reduction method. From the simulator outputs, we then define the difference function
$$ \Delta y_{HF-LF}(\bm{x}_{HF}) = y_{HF}(\bm{x}_{HF}) - y_{LF}(\bm{x}_{HF}),$$
perform fPCA on $\Delta y_{HF-LF}$, and fit a second surrogate model. We then sum surrogate predictions $\hat{y}_{LF}$ of the low-fidelity model and $\Delta \hat{y}_{HF-LF}$ of the difference function to form predictions
$$ \hat{y}_{HF}(\bm{x}) = \hat{y}_{LF}(\bm{x}) + \Delta \hat{y}_{HF-LF} (\bm{x}).$$
of the high-fidelity model.

\subsection{Fusion}
For the fusion approach, we follow the methodology described in \cite{benamara2017multi-fidelity-c96}. In this approach, we seek an enhanced basis that spans both the HF and LF data. We then project the LF functions onto the enhanced basis and fit a surrogate model to the coefficients. We then train a second surrogate model to model the difference in the coefficients from the HF and LF data on the enhanced basis.

In constructing a joint basis for the HF and LF functions, we first seek to find an orthonormal basis $\Phi$ for the data spanned by $y_{HF}$. Denoting $\mu_y$ as the mean function we use the standard $\mathbb{L}_2$ distance to compute the projection error
$$J(\mu_y, \Phi) = \frac{1}{2}\sum_{i=1}^{N_{HF}}\|y_i - \mu_y - \Phi\Phi^\top(y_i-\mu_y)\|^2.$$

Here $N_{HF}$ is the number of HF functions and to simplify the notation, we drop the HF on $y$. Additionally, $y_i$ denotes the $i^{th}$ HF function. We find this basis using the standard QR decomposition on the mean-centered HF functions,
$$ Y = [Q_1 | Q_2] R,$$
where we choose $\Phi = Q_1$.

We then seek to complete the joint HF-LF basis in the subspace that is orthogonal to the HF basis $\Phi$. We will denote $N_{LF}$ as the number of LF functions and as stated earlier, we assume $N >> n$. \cite{benamara2017multi-fidelity-c96} uses the following global cost function to find the optimal complete basis $\Psi$:

$$ J(\mu_\upsilon, \Psi) = \frac{1}{2}\sum_{i=1}^{N_{HF}}\|y_i - \mu_y - \Psi\Psi^\top(y_i-\mu_\upsilon)\|^2 + \frac{1}{2}\sum_{i=1}^{N_{LF}}\|\upsilon_i - \mu_y - \Psi\Psi^\top(\upsilon_i-\mu_\upsilon)\|^2,$$

where $\upsilon_i$ denotes the $i^{th}$ LF function and we define $\Psi = [\Phi | \Xi]$. The mean vector of the LF functions is defined as $\mu_\upsilon = \mu_y + d$. With $\Phi$ and $\Xi$ being orthonormal, the cost function reduces to

$$ J(\mu_\upsilon, \Psi) = \frac{N_{HF}}{2}d^\top (I-\Xi\Xi^\top) d + \frac{1}{2}\sum_{i=1}^{N_{LF}}\|(z^\perp_i - d)^\top (I - \Xi\Xi^\top)(z^\perp_i-d   )\|^2,$$
where $z_i^\perp = (I - \Phi\Phi^\top)z_i$. The solution for $d$ is 
$$d = \frac{N_{LF}}{N_{HF}+N_{LF}} \mu_{\upsilon^\perp},$$
where $\mu_{\upsilon^\perp}$ is the mean function of the LF functions projected on $\Phi$.

By centering $\upsilon_i^\perp$ with $d_{opt}$ and since the basis being is orthonormal, i.e., $\Phi^\top \Xi = 0$, we can express the cost function as 
$$ J(\Xi) = \frac{1}{2} \sum_{i=1}^{N_{LF}} u_i^\top (I-\Xi\Xi^\top) u_i,$$
where $u = \upsilon^\perp - d_{opt}$.  $\Xi$ is found using SVD which leads to the optimal mean $\mu_\upsilon = \mu_y + d_{opt}$ and $\Psi = [\Phi | \Xi]$.

We then project the LF functions on $\Psi$ and fit an emulator to the coefficients. We then compute compute the projection of the HF functions on $\Psi$ to get the coefficients $\eta_{HF}$ and then predict the coefficients ($\hat{\eta}_{HF})$ for the sample points at $\bm{x}_{HF}$. From the predicted coefficients we compute the residual 
$$ r = \eta_{HF} - \hat{\eta}_{HF}.$$
We then fit an emulator to predict the residuals, $y_{r}(\bm{x}_{HF})$. Prediction of a HF sample at $\bm{x}$ is then obtained with
$$y_{HF}(\bm{x}) = y_{LF}(\bm{x}) + y_{r} (\bm{x}).$$

\section{Results}
\label{sec:results}
\subsection{Simulation}
\label{sec:simulation}
We consider a simulation that comes from 2-dimensional design space, $\mathcal{D} \in \mathbb{R}^2$. We consider the simulation setup from \cite{benamara2017multi-fidelity-c96}, which is as follows:
The input parameter space is  $\mathcal{D} = [4,6]\times[10,14]$. The LF function is defined as
$$y_{LF}(t,x) = \frac{1}{2} \left((6t-2)^2\sin(x_2t-4)\right)+10\left(t-\frac{1}{2}\right)-x_1,$$
and the HF function is assumed as
$$y_{HF}(t,x) = \frac{1}{2} \left((6t-2)^2\sin(x_2t-4)\right)+\sin(10\cos(x_1t)).$$

\begin{figure}[htbp]
    \centering
    \includegraphics[width=1\linewidth]{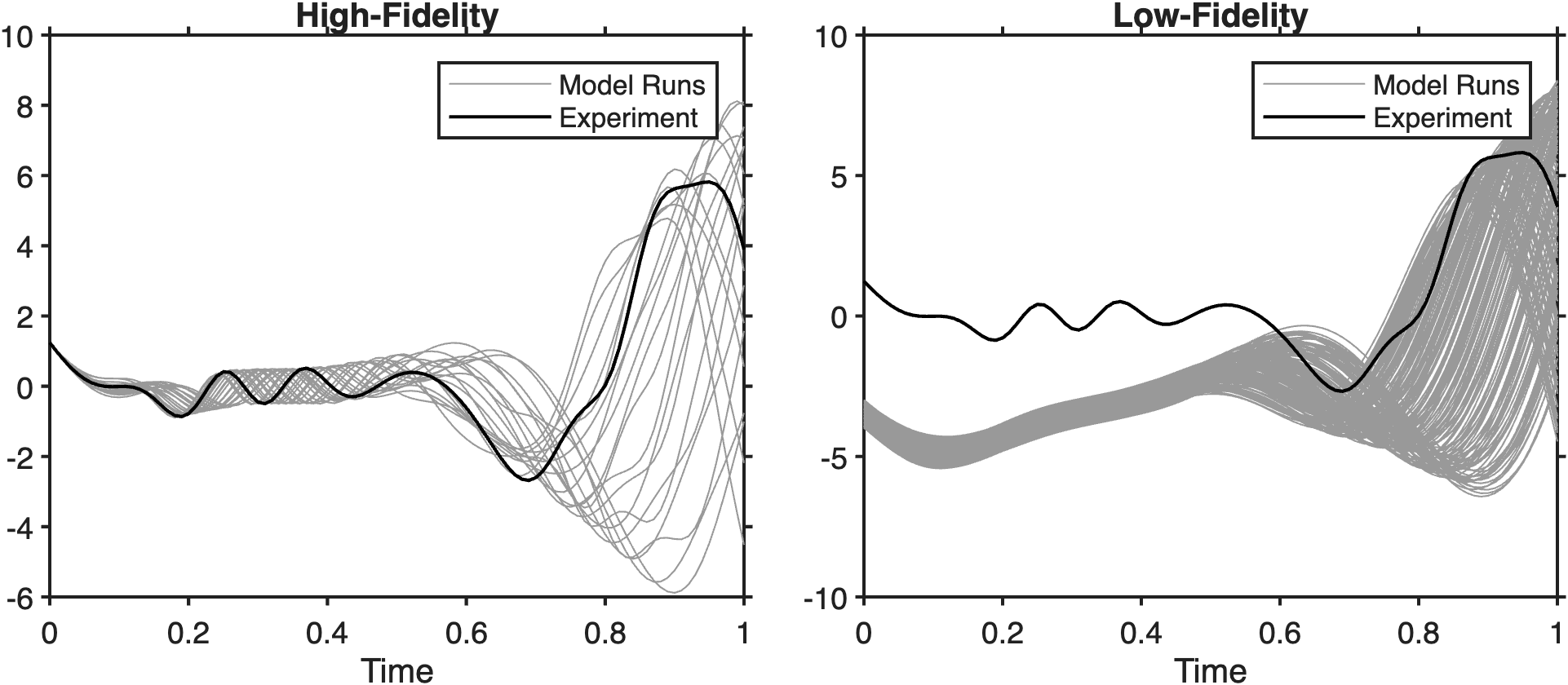}   
    \caption{Simulated dataset. Left: $n=20$ high-fidelity (HF) realizations. Right: $N=150$ low-fidelity (LF) samples. The solid black line corresponds to the experimental data.}
    \label{fig:toy_data}
\end{figure}

Figure~\ref{fig:toy_data} presents a set of simulations (model runs) from the HF (left panel)  and LF (right panel) functions respectively for set of inputs sampled via a latin-hypercube design. For the HF dataset, the number of samples is $n=20$ and for the LF dataset the number of samples is $N=150$. An experimental function from the HF case is shown as a black solid line which corresponds to the input parameters $\theta=[x_1,x_2]=[5.608, 12.987].$

\begin{figure}[htbp]
    \centering
    \includegraphics[width=1\linewidth]{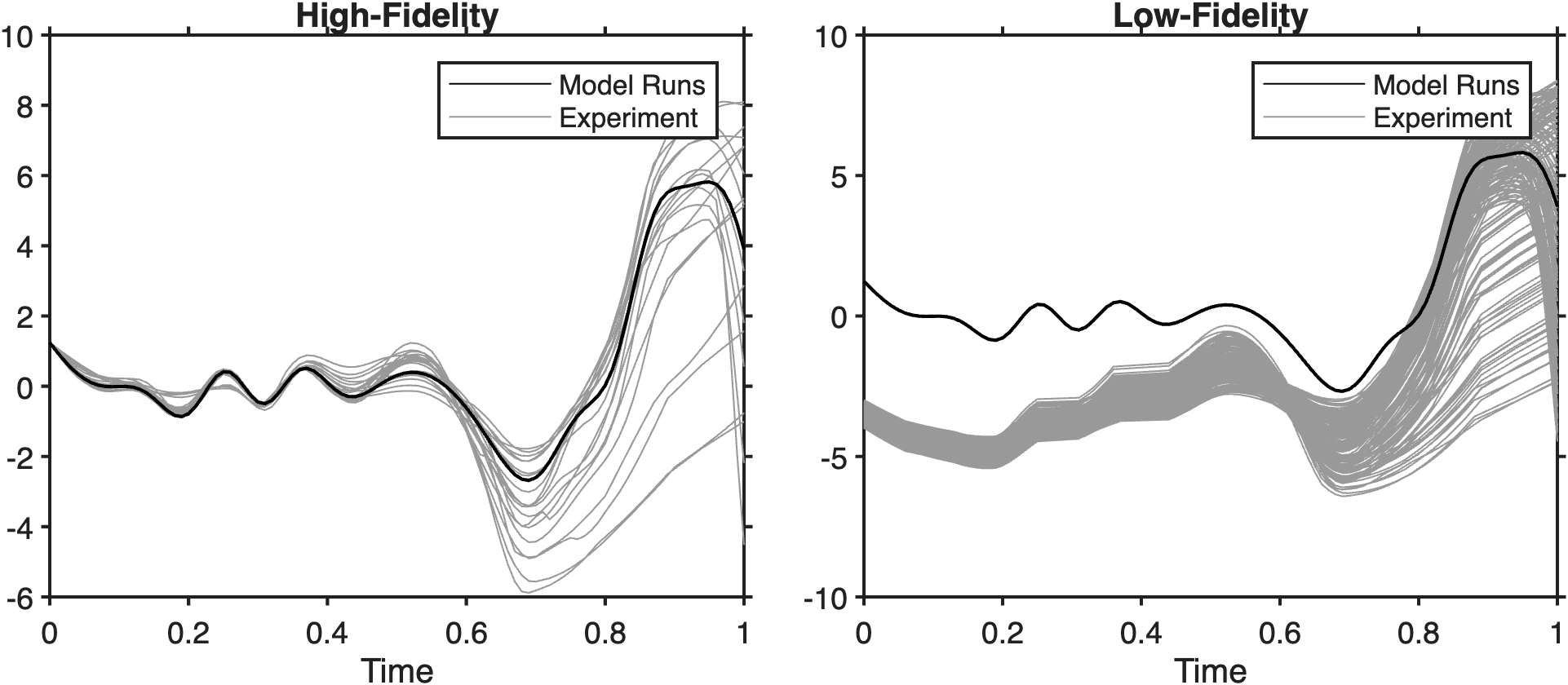}
    \caption{Aligned simulated dataset. Left: high-fidelity (HF) realizations. Right: low-fidelity (LF) samples. The solid black line corresponds to the experimental data.  }
    \label{fig:toy_data_aligned}
\end{figure}

\begin{figure}[htbp]
    \centering
    \includegraphics[width=1\linewidth]{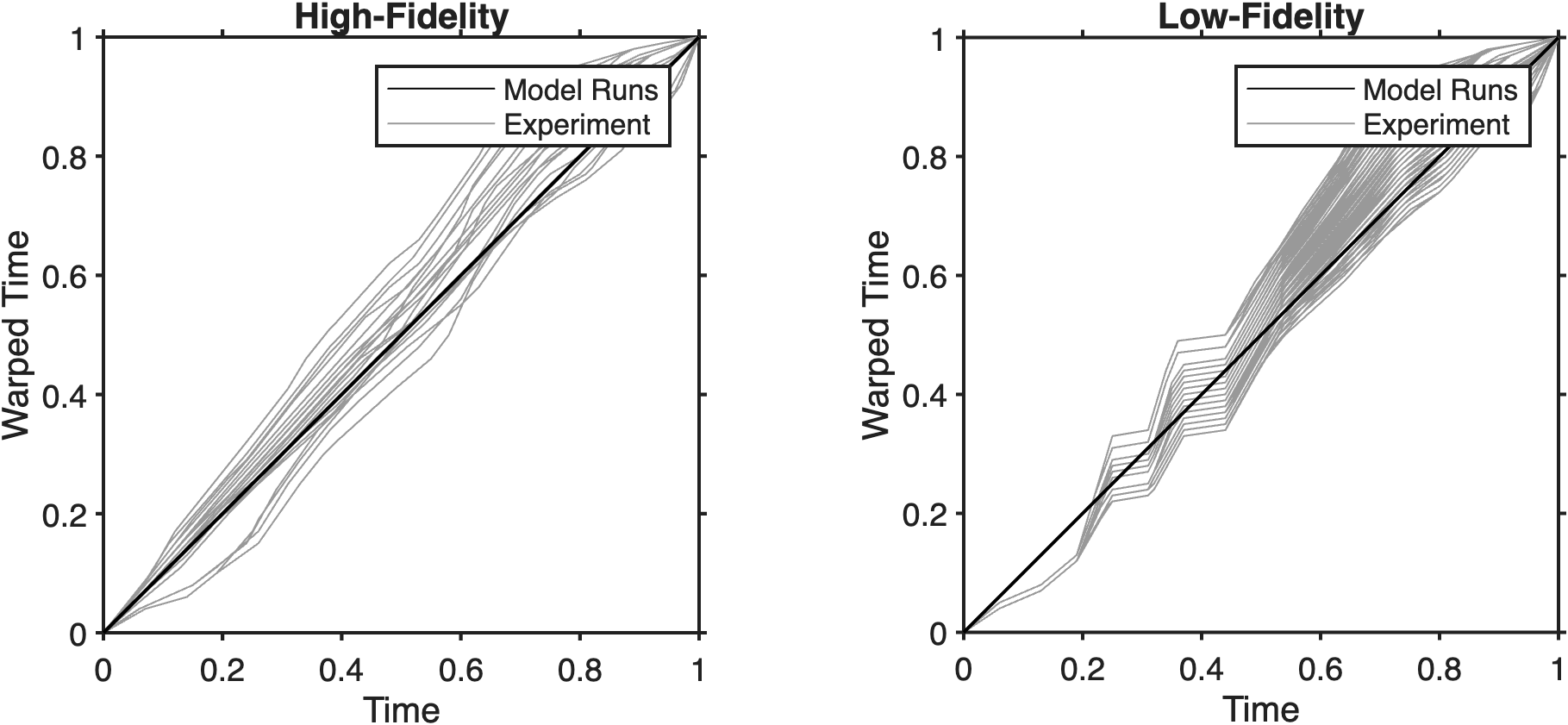}
    \caption{Warping functions for simulated dataset..Left: high-fidelity (HF) realizations. Right: low-fidelity (LF) samples. The solid black line corresponds to the experimental data.   }
    \label{fig:toy_data_warp}
\end{figure}

For elastic calibration, we align the simulation data to the experiment following \cite{francom:2025} as described in \ref{sec:align} and \ref{sec:alignment}. Figure~\ref{fig:toy_data_aligned} presents the aligned simulations to the experimental function for each of the HF and LF cases. The corresponding warping functions are shown in Figure~\ref{fig:toy_data_warp}. Note that when the simulated curves are aligned to the experiment data, the corresponding warping function is the identity function and its log-derivative representation is equal to zero.

We trained the emulators for both the mapping and fusion approaches using Bayesian Adaptive Spline Surfaces (BASS), \cite{francom2018bass}. BASS is an effective Bayesian regression model that has demonstrated high predictive accuracy and reliable uncertainty quantification across a wide range of complex physics-based applications, and, unlike Gaussian processes, scales well to moderate-to-large sample sizes, \cite{rumsey2025all}.  This is important for the calibration example presented in the next section. For the mapping approach, the aligned function emulator for $y_{LF}$ achieved an $R^2 = 0.95$, while the corresponding emulator for the log-derivative representation achieved an $R^2=0.97$. The aligned function emulator for $\Delta y_{HF-LF}$ achieved an $R^2=0.72$, and the emulator for the log-derivative representation achieved an $R^2=0.51$. The lower $R^2$ values for these emulators reflect the greater complexity of the differential space they represent. For the fusion approach, the emulator for the $y_{LF}$ aligned functions and for $y_4$ achieved a value of $R^2=0.99$. The emulator for the log-derivative representations achieved an $R^2=0.90$ and for $y_r$ achieved a value of $R^2=0.79$.

\begin{figure}
    \centering
    \includegraphics[width=.7\linewidth]{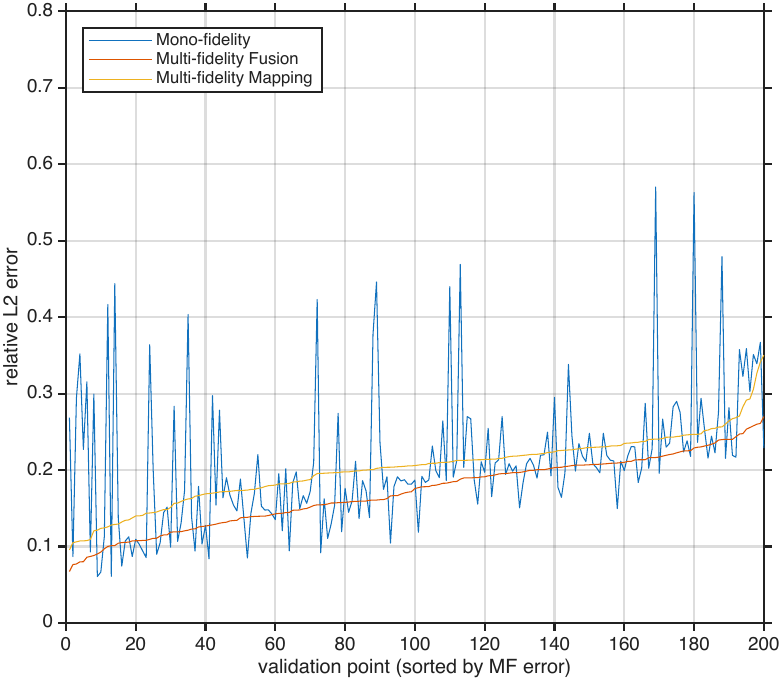}
    \caption{Comparison of mono-fidelity and multi-fidelity approaches with the Fusion approach and mapping approach using leave-one-out cross-validation.}
    \label{fig:emu_comparison}
\end{figure}

We evaluated both approaches using a leave-one-out cross-validation procedure.
In each fold, one input-response pair of the HF data was held out and treated as the synthetic observed experimental data, while the remaining pairs were used as simulation data to train the surrogate model and perform calibration. Figure~\ref{fig:emu_comparison} compares the multi-fidelity (MF) and standard mono-fidelity approaches, using only the HF samples for the latter, for both the fusion strategy (orange curve) and the mapping strategy (yellow curve). The curves are ordered by MF error, with the mono-fidelity results shown in blue and the MF results in orange and yellow. Error is measured as the relative $\mathbb{L}^2$ distance between the true function and the predicted function, after composing the predicted aligned function with the predicted warping function to reconstruct the prediction in the original data space. On the x-axis the curves are sorted by the MF error of the fusion approach. For both approaches, fusion or mapping the MF methods yields an average error that is comparable to or lower than that of the mono-fidelity method. Comparing the two MF strategies, the fusion approach achieves the lower average relative $\mathbb{L}^2$ error.

\begin{figure}[htbp]
    \centering
    \includegraphics[width=.7\linewidth]{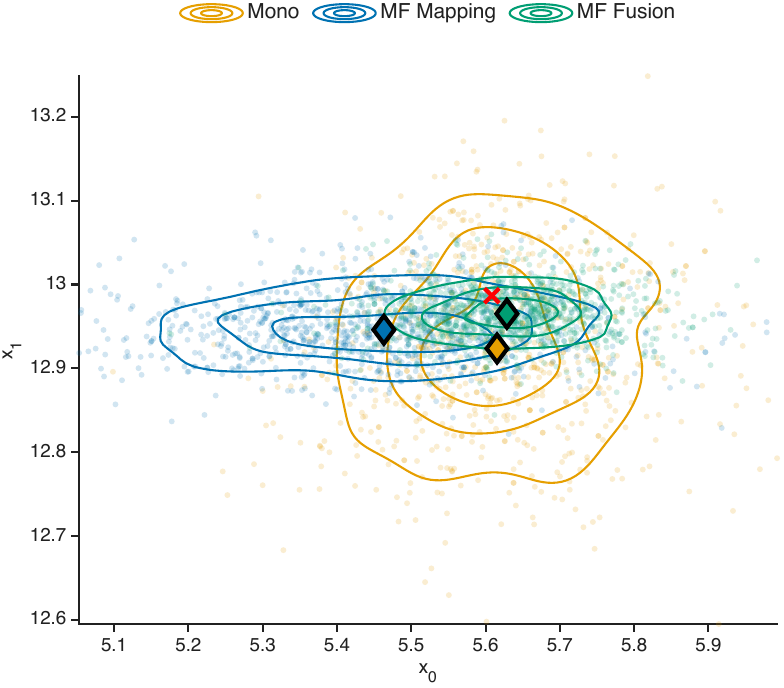}
    \caption{Posterior distributions for the mono fidelity and two multi-fidelity approaches. The orange contours represent the mono fidelity while the blue and green contours represent the posterior for the MF mapping and the MF fusion approache.}
    \label{fig:pair_all}
\end{figure}

Furthermore, we produced a Bayesian model calibration to the experimental function using the two MF surrogate methods as  described in \cite{francom:2025} and in \ref{sec:ebcalibration} and a mono fidelity approach that only relies on the MF samples.  Figure~\ref{fig:pair_all} presents contour plots for the bivariate posterior distribution of the parameters, $x_0$, and $x_1$ based on the MCMC samples from the calibrations. The orange contours represents the posterior distribution from the mono-fidelity approach using the HF samples. The blue contours represent the posterior distribution for the mapping MF approach, while the green contours the posterior for the Fusion MF approach. The diamonds are the posterior median values and the X in red is the true point that generated the data. From this figure, we notice that both of the MF approaches reduce the spread of the posterior distribution with the Fusion approach having lower uncertainty and the best coverage of the true parameter values. The reduction in uncertainty comes from incorporating samples from the LF computational model and using a proper metric within the elastic FDA framework.

\subsection{Tin Calibration}
\label{sec:tin}
\begin{figure}[htbp]
\centering
\includegraphics[width=0.65\textwidth]{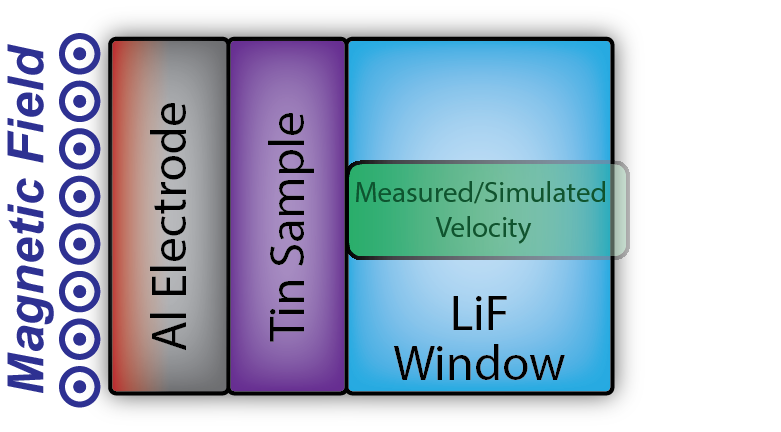}
\caption{Set up for the Thor mini-Z-machine experiments.}
\label{fig:zmachine}
\end{figure}

In the context of an equation-of-state (EoS) for tin generated from pulsed magnetic fields \citep{brown2023}, we estimate the parameters describing the compressibility (relationship between pressure and density) to understand better how materials compress to extreme pressures, via the MF and HF approaches described in Section \ref{sec:mf}.
In this case, the experimental data was generated using Sandia National Laboratories' mini-Z-machine called Thor, which is a pulsed power driver that can deliver massive electric currents over short time scales. These currents were forced to flow along an aluminum (Al) panel, producing a large magnetic pressure which drives a time-dependent stress wave (impulse) into the system. 
Tin samples and transparent lithium fluoride (LiF) windows were glued to the panel such that the stress wave propagates sequentially through each of these materials. A graphical depiction of the experiments is shown in Figure~\ref{fig:zmachine}.

The tin experiments are designed to investigate both the strength and the phase-transition kinetics of a two-phase system consisting of the ambient beta phase and the high-pressure gamma phase. The parameters $b_{yfac}$ and $g_{yfac}$ relate to the yield strength of the respective phases. In addition, each phase transformation is governed by two kinetics parameters, $(B,\nu)$. Specifically, $B_{bg}$ and $\nu_{bg}$ govern the forward beta-gamma transformation, while $B_{gb}$ and $\nu_{gb}$ govern the reverse transition. Together with the remaining material parameters, this yields a total of $p=11$ input parameters to be considered for surrogate-model construction and calibration. Additional discussion of these material parameters is provided in \cite{met12111844}, and a scalar Bayesian calibration for related data is presented in \cite{tin_llnl}.

\begin{figure}[htbp]
    \centering
    \includegraphics[width=1\linewidth]{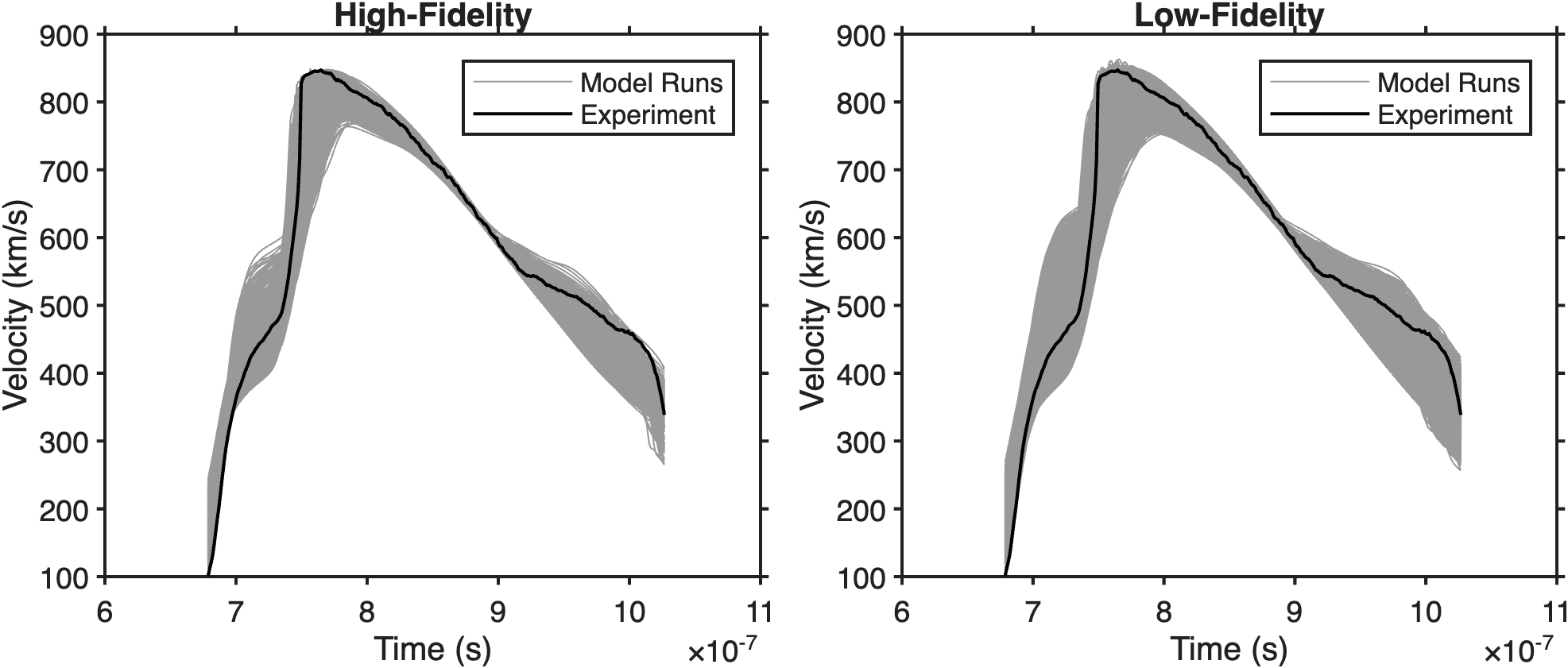}
    \caption{Original Tin Z-machine experiment (black) and high-fidelity (left) and low-fidelity (right) computer model runs.}
    \label{fig:tin_f}
\end{figure}

Figure~\ref{fig:tin_f} presents the velocity curves resulting from the experiments for Tin at the Z-machine with the corresponding HF and LF computer runs shown as the gray lines. There are $n=500$ HF simulations and $5500$ LF simulations. The difference in fidelity is a result of the mesh resolution and that the physics code is considered numerically converged for approximately a 1 micron mesh size for the HF simulations. The LF simulations are done at a mesh size of 10 microns and run computationally faster than the HF simulations by a factor of 25. The LF mesh size is under-resolved and as a consequence, becomes heavily influenced by the artificial viscosity - effectively a numerical tuning parameter that influences the shape (rise time) of the wave. This introduces a known simulator error where the shape of the curves are about right but the fine details are wrong.

\begin{figure}[htbp]
    \centering
    \includegraphics[width=1\linewidth]{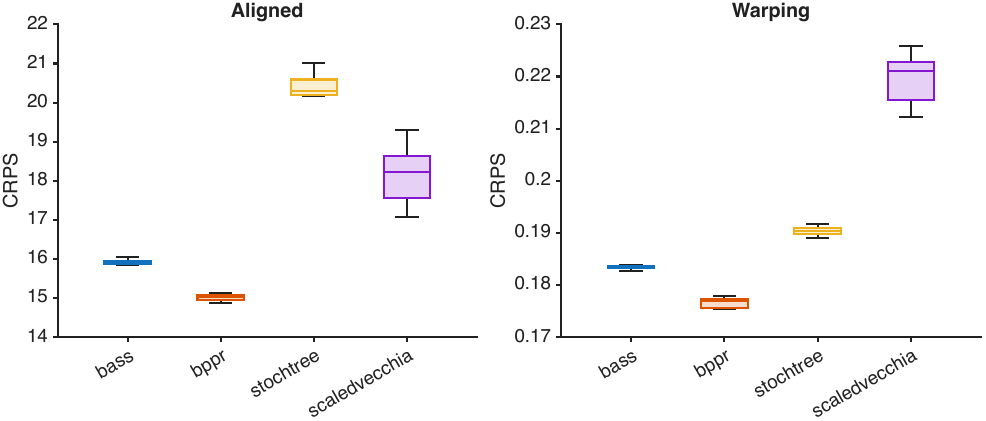}
    \caption{Boxplots of Continuous Ranked Probability Score (CRPS) for the aligned (left) and warping functions (right). The boxplots represent the CRPS computed over a 5-fold cross-validation}
    \label{fig:tin_crps}
\end{figure}

As in the simulation examples of \ref{sec:simulation} , the computer model output was aligned to the experimental data, and an emulator or surrogate was fitted to the aligned computer model output and its corresponding log-derivative representation. To determine the best emulator, we ran a 5-fold cross-validation on both the HF and LF data for both the aligned functions and the log-derivative representations. We considered and compared 4 emulators: BASS (\cite{francom2018bass}), Bayesian Project Pursuit Regression (BayesPPR \cite{collins2024bayesian}), Bayesian Adaptive Regression Trees (BART \cite{chipman2010bart}), and Scaled Vecchia Gaussian Process Regression (ScaledVecchia \cite{katzfuss2022scaled}). These emulators represent a broad class of methods capable of efficiently handling thousands of observations. Additionally, each of these emulators performed well in an extensive emulator comparison provided by \cite{rumsey2025all}. We utilized the corresponding R packages for BASS (\cite{francom2019JSS}), BayesPPR (\cite{bayesppr}) and ScaledVechhia while for BART we used the implementation in StochTree (\cite{herren2025stochtree}). Figure~\ref{fig:tin_crps} presents boxplots of the Continuous Ranked Probability Score (CRPS) for each of the 4 multi-fidelity emulators for the aligned (left) and warping functions (right). Given that BPPR has the lowest CRPS, we chose the BPPR surrogate model to compare the MF and HF approach to calibration of the tin data.

\begin{figure}[htpb]
    \centering
    \includegraphics[width=1\linewidth]{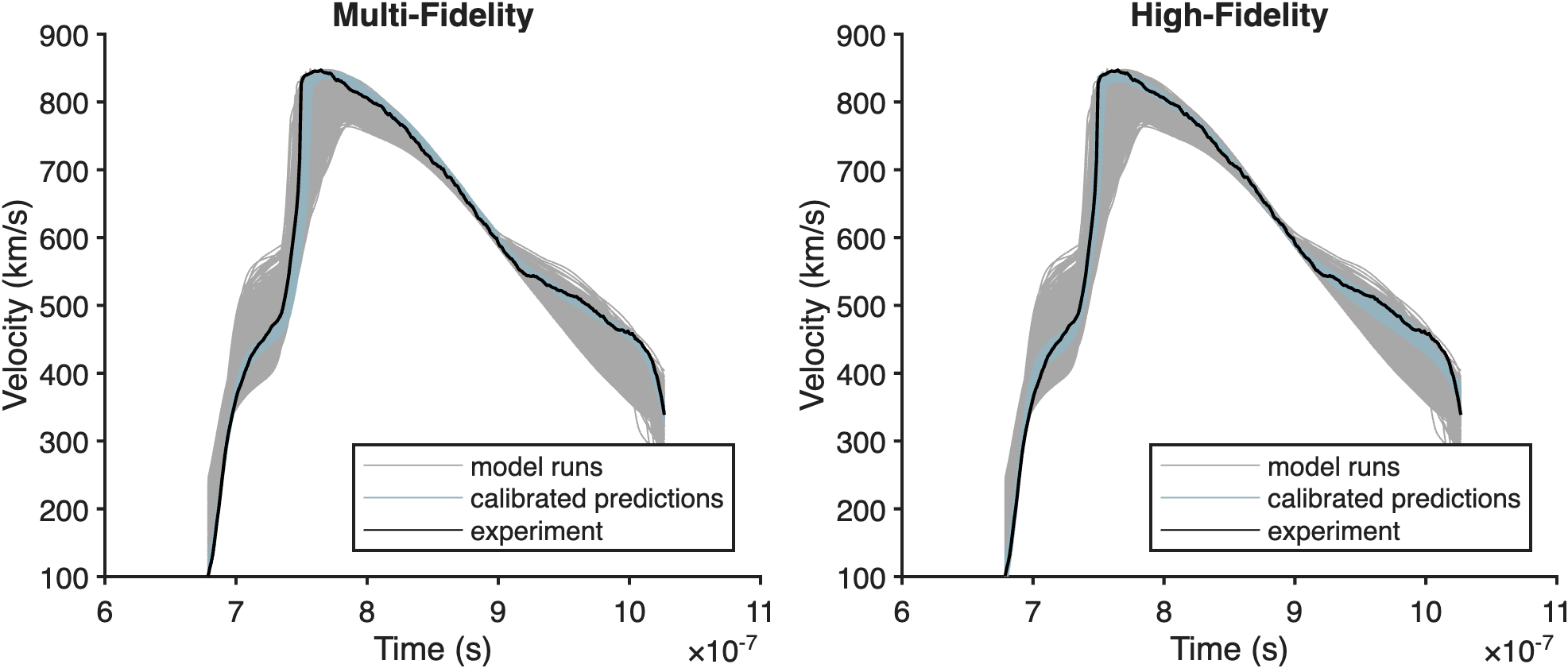}
    \caption{Experiment velocities of tin shown in black with 95\% posterior prediction intervals from the elastic functional Bayesian calibration shown in blue. Model runs are shown in grey. The left panel corresponds to the multi-fidelity approach and the right panel considers high-fidelity samples only.}
    \label{fig:tin_cal}
\end{figure}

We then performed a modular elastic Bayesian model calibration utilizing the formulation in \ref{sec:ebcalibration}. With the increased performance of the fusion approach over the mapping approach show in the simulated data set in \ref{sec:simulation}, we employ for this application. Figure~\ref{fig:tin_cal} presents post-calibration prediction results for the tin experiment. 
The black curves shown in the figure correspond to the experimental velocity curves and the gray curves are the high-fidelity computer model runs.
The shaded colored regions are the 95\% prediction intervals that result from the elastic functional Bayesian calibration. The prediction intervals exhibit good agreement with each experimental curve. The left panel shows results from the multi-fidelity approach, while the right panel shows results from using the HF samples only approach. Utilizing the LF samples improves coverage of the shock region and late time of the signal, compared to using only the HF samples alone. 

\begin{figure}[htpb]
    \centering
    \includegraphics[width=1\linewidth]{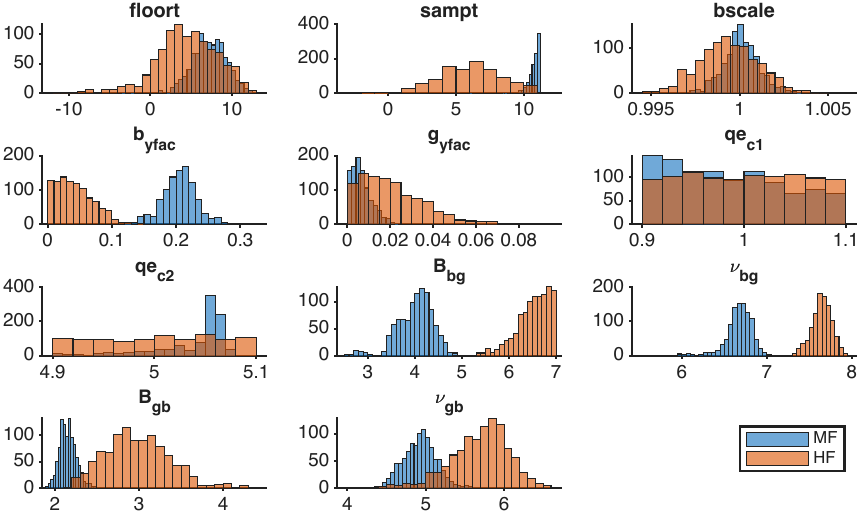}
    \caption{Histograms of samples from the posterior distribution of the EoS parameters for tin. Multi-fidelity histograms in blue and high-fidelity in orange .}
    \label{fig:tin_histograms}
\end{figure}

Furthermore, Figure~\ref{fig:tin_histograms} presents the histograms of the posterior distribution samples from the calibrated EoS parameters for tin with the elastic approach. The histograms resulting from the multi-fidelity approach are shown in blue and the histograms from the high-fidelity only case are shown in orange. When utilizing the LF samples in our multi-fidelity framework, we notice a tighter posterior distributions for each of the parameters compared to the ones resulting from the elastic calibration with HF samples only. The increased number of samples help increase the precision of the calibration, as was also demonstrated in the simulation example of \ref{sec:simulation}. These results are comparable with recent work in \cite{prime:2026} where a similar inference for  $g_{yfac}$ was achieved. 

\section{Conclusion and Future Work}
\label{sec:discussion}

Calibration of computer models with functional response is complicated by two largely independent difficulties. The first is representational: when the model's parameters control the timing or location of features as well as their magnitude, dimension reduction schemes that assume amplitude-only variation produce inefficient bases, degraded emulators, and calibration likelihoods built on a metric that is not a proper distance. The second is economic: characterizing a posterior distribution requires many simulator evaluations, and converged high-fidelity simulators are often too expensive to provide. The elastic calibration framework of \cite{francom:2025} addresses the first difficulty and the multi-fidelity surrogate literature addresses the second, but until now, the two approaches have been pursued separately. In particular, the multi-fidelity constructions have inherited the amplitude-only assumption of the majority of calibration methods with the notable exception of \cite{francom:2025}.

The main goal of this paper is to demonstrate elastic model calibration in a multi-fidelity framework. By aligning both the high- and low-fidelity ensembles to a common reference before any surrogate model is constructed, the misaligned functional output of each fidelity level is separated into an aligned function and a warping function, and the standard multi-fidelity constructions can then be applied to each component independently. We demonstrated both of the dominant strategies in this setting: the mapping approach, which emulates the low-fidelity response and a correction toward the high-fidelity response, and the fusion approach of \cite{benamara2017multi-fidelity-c96}, which constructs an enhanced basis spanning both ensembles and emulates the coefficients together with a residual correction. Representing the phase component through the log-derivative transform of \cite{ma2024stochastic} rather than through a tangent-space approximation is what makes this possible. The resulting space is a closed linear subspace of $\ltwo$, so the differencing and projection operations on which both strategies rely are well defined for the phase component and always return valid warping functions. The phase and amplitude aspects of the experiments and computer model output receive the same treatment.

The two-dimensional example in \ref{sec:simulation} demonstrated that with $N = 150$ low-fidelity runs supplementing only $n = 20$ high-fidelity runs, both elastic multi-fidelity strategies matched or improved on the leave-one-out predictive accuracy of a mono-fidelity elastic emulator trained on the high-fidelity data alone. Also, the fusion approach achieved the lowest relative $\ltwo$ error on average. These result carried through to the calibration itself: both multi-fidelity posteriors were tighter than the mono-fidelity posterior, and the fusion posterior had the best coverage and a posterior median closest to the true parameter values. The application to the tin equation-of-state calibration problem showed that the apparent benefits with synthetic data persists in a realistic setting with $p = 11$ inputs, where the fidelity levels differ in mesh resolution rather than in analytic form. Therefore, the low-fidelity ensemble improved coverage of the shock region and late time of the signal and tightened the posterior distribution of each calibrated parameter. These results indicate that inexpensive low-fidelity functional data, provided that it is properly aligned before used, can meaningfully improve calibration when the high-fidelity budget is limited. 

Several future directions remain open, the approaches employed here used two fidelity levels. A natural next step is to extend the elastic construction to a hierarchy with three or more levels, similar to multilevel methods. Additionally, it is important to address how to optimally allocate a fixed computational budget across these levels when the target quantity is a phase-amplitude decomposition instead of a scalar.


A second set of concerns are related to the alignment itself. As in \cite{francom:2025}, the warping functions are obtained by optimization and then treated as fixed, so uncertainty introduced by the warping decomposition is not propagated through the calibration. This uncertainty is small in the low-noise settings considered here. However a more complete treatment, perhaps through the modular Bayesian techniques of \cite{liu2009modularization} and \cite{plummer2015cuts} or through a fully Bayesian alignment \citep{cheng2016bayesian, lu2017bayesian}, would be valuable, particularly for experimental data with appreciable measurement noise. The assumption that the aligned functions $\tilde y$ and the log-derivative representations $\bm h$ are modeled independently likewise warrants further assessment, and is potentially more consequential in the multi-fidelity setting, where a correlation between the amplitude and phase corrections across fidelity levels would be discarded by the independent treatment. The choice of alignment reference merits attention as well. Aligning both LF and HF ensembles to the experimental curve is natural when calibration is the goal, but aligning instead to a noiseless common reference drawn from the model runs may prove more stable. The sensitivity of the multi-fidelity constructions to this alternative choice has not been characterized.

Lastly, model discrepancy is considered to be negligible in the examples presented in Section \ref{sec:results}. The elastic framework accommodates discrepancy in both amplitude and phase space, and the multi-fidelity constructions do not preclude it. However, a discrepancy model that is estimated jointly with a multi-fidelity surrogate may raise identifiability issues, since low-fidelity bias and model form error both manifest as systematic departures from the high-fidelity response. Disentangling the two is an important problem for extensions of our proposed framework, in which the low-fidelity model is a genuine physical simplification rather than a coarsened discretization.

\section{Funding}
This paper describes objective technical results and analysis. Any subjective views or opinions that might be expressed in the paper do not necessarily represent the views of the U.S. Department of Energy or the United States Government.

This article has been authored by an employee of National Technology \& Engineering Solutions of Sandia, LLC under Contract No. DE-NA0003525 with the U.S. Department of Energy (DOE). The employee owns all right, title and interest in and to the article and is solely responsible for its contents. The United States Government retains and the publisher, by accepting the article for publication, acknowledges that the United States Government retains a non-exclusive, paid-up, irrevocable, world-wide license to publish or reproduce the published form of this article or allow others to do so, for United States Government purposes. The DOE will provide public access to these results of federally sponsored research in accordance with the DOE Public Access Plan https://www.energy.gov/downloads/doe-public-access-plan

\bibliographystyle{plainnat}
\bibliography{refs}

@article{bayarri2007framework,
  author  = {Bayarri, M. J. and Berger, J. O. and Paulo, R. and Sacks, J. and Cafeo, J. A. and Cavendish, J. and Lin, C. and Tu, J.},
  journal = {Technometrics},
  number  = {2},
  title   = {A framework for validation of computer models},
  volume  = {49},
  year    = {2007}
}

@article{benamara2017multi-fidelity-c96,
  year    = {2017},
  title   = {Multi-fidelity {POD} surrogate-assisted optimization: Concept and aero-design study},
  author  = {T. Benamara and P. Breitkopf and I. Lepot and C. Sainvitu and P. Villon},
  journal = {Structural and Multidisciplinary Optimization},
  pages   = {1387–1412},
  volume  = {56}
}

@article{Bomarito2022,
  author  = {G. F. Bomarito and P. E. Leser and J. E. Warner and W. P. Leser},
  title   = {On the optimization of approximate control variates with parametrically defined estimators},
  journal = {Journal of Computational Physics},
  volume  = {451},
  pages   = {110882},
  year    = {2022},
  doi     = {https://doi.org/10.1016/j.jcp.2021.110882}
}

@article{brown2023,
  author  = {J. L. Brown and J. P. Davis and J. D. Tucker and G. Huerta and K. W. Shuler},
  title   = {Quantifying Uncertainty in Analysis of Shockless Dynamic Compression Experiments on Platinum. II. Bayesian Model Calibration},
  journal = {Journal of Applied Physics},
  volume  = {134},
  number  = {23},
  pages   = {235902},
  year    = {2023},
  month   = {12},
  issn    = {0021-8979},
  doi     = {10.1063/5.0173652}
}

@article{BRUNEL2025117577,
  title    = {A survey on multi-fidelity surrogates for simulators with functional outputs: Unified framework and benchmark},
  journal  = {Computer Methods in Applied Mechanics and Engineering},
  volume   = {435},
  pages    = {117577},
  year     = {2025},
  issn     = {0045-7825},
  author   = {L. Brunel and M. Balesdent and L. Brevault and R. {Le Riche} and B. Sudret}
}

@article{bryn2014,
  author        = {Brynjarsd{\'o}ttir, J. and O'Hagan, A.},
  journal       = {Inverse Problems},
  number        = {11},
  pages         = {114007},
  title         = {{Learning about physical parameters: The importance of model discrepancy}},
  volume        = {30},
  year          = {2014}
}

@article{cheng2016bayesian,
  author    = {Cheng, W. and Dryden, I. L. and Huang, X. and others},
  journal   = {Bayesian Analysis},
  number    = {2},
  pages     = {447--475},
  publisher = {International Society for Bayesian Analysis},
  title     = {Bayesian registration of functions and curves},
  volume    = {11},
  year      = {2016}
}

@article{chipman2010bart,
  author  = {Chipman, H. A. and George, E. I. and McCulloch, R. E.},
  title   = {{BART}: {B}ayesian Additive Regression Trees},
  journal = {The Annals of Applied Statistics},
  year    = {2010},
  volume  = {4},
  number  = {1},
  pages   = {266--298},
  doi     = {10.1214/09-AOAS285}
}

@article{collins2024bayesian,
  title     = {Bayesian projection pursuit regression},
  author    = {G. Collins and D. Francom and K. Rumsey},
  journal   = {Statistics and Computing},
  volume    = {34},
  number    = {1},
  pages     = {29},
  year      = {2024},
  publisher = {Springer}
}

@inproceedings{Eld04,
  author    = {Eldred, M.~S. and Giunta, A.~A. and Collis, S.~S.},
  title     = {Second-Order Corrections for Surrogate-Based Optimization with Model Hierarchies},
  booktitle = {Proceedings of the 10th AIAA/ISSMO Multidisciplinary Analysis and Optimization Conference},
  year      = {2004},
  address   = {Albany, NY,},
  month     = {Aug. 30--Sept. 1,},
  note      = {AIAA Paper 2004-4457}
}

@inproceedings{Eld06b,
  author    = {Eldred, M.~S. and Dunlavy, D.~M.},
  title     = {Formulations for Surrogate-Based Optimization with Data Fit, Multifidelity, and Reduced-Order Models},
  booktitle = {Proceedings of the 11th AIAA/ISSMO Multidisciplinary Analysis and Optimization Conference},
  year      = 2006,
  month     = {September~6--8},
  series    = {AIAA-2006-7117},
  address   = {Portsmouth,~VA}
}

@inbook{Eldred2017,
  author    = {M. S. Eldred
               and L. W. T. Ng
               and M. F. Barone
               and S. P. Domino},
  title     = {Multifidelity Uncertainty Quantification Using Spectral Stochastic Discrepancy Models},
  booktitle = {Handbook of Uncertainty Quantification},
  year      = {2017},
  publisher = {Springer International Publishing},
  address   = {Cham},
  pages     = {991--1036},
  isbn      = {978-3-319-12385-1},
  doi       = {10.1007/978-3-319-12385-1_25}
}

@article{francom:2025,
  author  = {D. Francom and J. D. Tucker and G. Huerta and K. Shuler and D. Ries},
  title   = {Elastic {B}ayesian Model Calibration},
  journal = {SIAM/ASA Journal on Uncertainty Quantification},
  volume  = {13},
  number  = {1},
  pages   = {195-227},
  year    = {2025},
  doi     = {10.1137/24M1644092}
}

@article{francom2018bass,
  author        = {D. Francom and B. Sanso and A. Kupresanin and G. Johannesson},
  journal       = {Statistica Sinica},
  number        = {2},
  pages         = {791-816},
  title         = {{Sensitivity analysis and emulation for functional data using Bayesian adaptive splines}},
  volume        = {28},
  year          = {2018}
}

@article{francom2019inferring,
  author        = {D. Francom and B. Sanso and V. Bulaevskaya and D. Lucas and M. Simpson},
  journal       = {Journal of the American Statistical Association},
  number        = {528},
  pages         = {1450-1465},
  title         = {{Inferring Atmospheric Release Characteristics in a Large Computer Experiment using Bayesian Adaptive Splines}},
  volume        = {114},
  year          = {2019}
}

@software{bayesppr,
  author  = {Collins, G. Q. and Tucker, J. D.},
  title   = {BayesPPR},
  url     = {https://cran.r-project.org/web/packages/BayesPPR/index.html},
  version = {0.1.0},
  year    = {2026},
  date    = {2026-05-18}
}

@article{francom2019JSS,
  title   = {{BASS: An R package for fitting and performing sensitivity analysis of Bayesian adaptive spline surfaces}},
  author  = {Francom, D. and Sanso, B.},
  journal = {Journal of Statistical Software},
  year    = {2020}
}

@article{francom2022landmark,
  author        = {D. Francom and B. Sanso and A. Kupresanin},
  journal       = {SIAM/ASA Journal on Uncertainty Quantification},
  number        = {1},
  pages         = {125-150},
  title         = {Landmark-Warped Emulators for Models with Misaligned Functional Response},
  volume        = {10},
  year          = {2022}
}

@inproceedings{geraci_multi-fidelity_2017,
  title     = {A multifidelity multilevel {Monte} {Carlo} method for uncertainty propagation in aerospace applications},
  isbn      = {978-1-62410-452-7},
  url       = {http://arc.aiaa.org/doi/10.2514/6.2017-1951},
  doi       = {10.2514/6.2017-1951},
  language  = {en},
  urldate   = {2019-10-04},
  booktitle = {19th {AIAA} {Non}-{Deterministic} {Approaches} {Conference}},
  publisher = {AIAA},
  author    = {G. Geraci and M.S. Eldred and G. Iaccarino},
  month     = jan,
  year      = {2017}
}

@article{GeraciCTR,
  title   = {A multi fidelity control variate approach for the multilevel Monte Carlo technique.},
  author  = {G. Geraci and G. Iaccarino and M. S. Eldred},
  year    = {2015},
  journal = {CTR Annual Research Briefs 2015},
  pages   = {169--181}
}

@article{Giles2008,
  author    = {M. B. Giles},
  issn      = {0030-364X},
  journal   = {Operations Research},
  month     = {June},
  number    = {3},
  pages     = {607--617},
  publisher = { INFORMS },
  title     = {{Multilevel Monte Carlo Path Simulation}},
  volume    = {56},
  year      = {2008}
}

@article{gorodetsky2019continuous,
  author  = {A. A. Gorodetsky and S. Karaman and Y. M. Marzouk},
  title   = {A continuous analogue of the tensor-train decomposition},
  journal = {Computer Methods in Applied Mechanics and Engineering},
  volume  = {347},
  pages   = {59--84},
  year    = {2019},
  issn    = {0045-7825},
  doi     = {10.1016/j.cma.2018.12.015}
}

@article{GORODETSKY2020109257,
  author  = {A. A. Gorodetsky and G. Geraci and M. S. Eldred and J. D. Jakeman},
  title   = {A generalized approximate control variate framework for multifidelity uncertainty quantification},
  journal = {Journal of Computational Physics},
  volume  = {408},
  pages   = {109257},
  year    = {2020},
  issn    = {0021-9991}
}

@article{gu2016parallel,
  author        = {M. Gu and J. O. Berger},
  journal       = {The Annals of Applied Statistics},
  number        = {3},
  pages         = {1317-1347},
  title         = {{Parallel Partial Gaussian Process Emulation for Computer Models with Massive Output}},
  volume        = {10},
  year          = {2016}
}

@article{happ2019,
  title   = {A general framework for multivariate functional principal component analysis of amplitude and phase variation},
  author  = {C. Happ and F. Scheipl and A. Gabriel and S. Greven},
  journal = {Stat},
  volume  = {8},
  number  = {1},
  pages   = {e220},
  year    = {2019},
  doi     = {10.1002/sta4.220}
}

@misc{herren2025stochtree,
  title         = {StochTree: BART-based modeling in R and Python},
  author        = {D. Herren and P. R. Hahn and J. S. Murray and C. M. Carvalho and J. He},
  year          = {2025},
  eprint        = {2512.12051},
  archiveprefix = {arXiv},
  primaryclass  = {stat.CO}
}

@article{higdon2004combining,
  author    = {Higdon, D. and Kennedy, M. and Cavendish, J. C. and Cafeo, J. A. and Ryne, R. D.},
  journal   = {SIAM Journal on Scientific Computing},
  number    = {2},
  pages     = {448--466},
  publisher = {SIAM},
  title     = {Combining field data and computer simulations for calibration and prediction},
  volume    = {26},
  year      = {2004}
}

@article{higdon2008computer,
  author  = {Higdon, D. and Gattiker, J. and Williams, B. and Rightley, M.},
  journal = {Journal of the American Statistical Association},
  number  = {482},
  title   = {Computer model calibration using high-dimensional output},
  volume  = {103},
  year    = {2008}
}

@book{horvath2012,
  author    = {Horvath, L. and Kokoszka, P.},
  publisher = {Springer},
  title     = {Inference for Functional Data with Applications},
  year      = {2012}
}

@article{katzfuss2022scaled,
  title     = {Scaled Vecchia approximation for fast computer-model emulation},
  author    = {Katzfuss, M. and Guinness, J. and Lawrence, E.},
  journal   = {SIAM/ASA Journal on Uncertainty Quantification},
  volume    = {10},
  number    = {2},
  pages     = {537--554},
  year      = {2022},
  publisher = {SIAM}
}

@article{kennedy2001bayesian,
  author        = {Kennedy, M. C. and O'Hagan, A.},
  journal       = {Journal of the Royal Statistical Society. Series B, Statistical Methodology},
  pages         = {425--464},
  title         = {Bayesian calibration of computer models},
  year          = {2001}
}

@article{kleiber2014model,
  author    = {Kleiber, W. and Sain, S. R. and Wiltberger, M. J.},
  journal   = {SIAM/ASA Journal on Uncertainty Quantification},
  number    = {1},
  pages     = {545--563},
  publisher = {SIAM},
  title     = {Model calibration via deformation},
  volume    = {2},
  year      = {2014}
}

@article{liu2009modularization,
  author    = {Liu, F. and Bayarri, M. J. and Berger, J. O. and others},
  journal   = {Bayesian Analysis},
  number    = {1},
  pages     = {119--150},
  publisher = {International Society for Bayesian Analysis},
  title     = {Modularization in {Bayesian} analysis, with emphasis on analysis of computer models},
  volume    = {4},
  year      = {2009}
}

@article{lu2017bayesian,
  author        = {Lu, Y. and Herbei, R. and Kurtek, S.},
  journal       = {Journal of Computational and Graphical Statistics},
  number        = {4},
  pages         = {894-904},
  publisher     = {Taylor \& Francis},
  title         = {{Bayesian Registration of Functions with a Gaussian Process Prior}},
  volume        = {26},
  year          = {2017}
}

@article{ma2024stochastic,
  title   = {A stochastic process representation for time warping functions},
  author  = {Y. Ma and X. Zhou and W. Wu},
  journal = {Computational Statistics \& Data Analysis},
  volume  = {194},
  pages   = {107941},
  year    = {2024},
  issn    = {0167-9473},
  doi     = {10.1016/j.csda.2024.107941}
}

@article{met12111844,
  author         = {N. R. Barton and D. J. Luscher and C. Battaile and J. L. Brown and M. Buechler and L. Burakovsky and S. Crockett and C. Greeff and A. E. Mattsson and M. B. Prime and W. J. Schill},
  title          = {A Multi-Phase Modeling Framework Suitable for Dynamic Applications},
  journal        = {Metals},
  volume         = {12},
  year           = {2022},
  number         = {11},
  article-number = {1844},
  issn           = {2075-4701},
  doi            = {10.3390/met12111844}
}

@article{Ng2014,
  title   = {Multifidelity approaches for optimization under uncertainty.},
  author  = {Ng., L. W. T. and Willcox, K. E.},
  year    = {2014},
  journal = {International Journal for numerical methods in Engineering},
  volume  = {100},
  number  = {10},
  pages   = {746--772}
}

@inproceedings{NgEldred2012,
  author    = {Ng, L.~W.~T. and Eldred, M.~S.},
  title     = {Multifidelity Uncertainty Quantification Using Nonintrusive Polynomial Chaos and Stochastic Collocation},
  booktitle = {Proceedings of the 14th AIAA Non-Deterministic Approaches Conference},
  series    = {AIAA-2012-1852},
  year      = {2012},
  month     = {April 23-26,},
  address   = {Honolulu, HI}
}

@article{peherstorfer2016optimal,
  title     = {Optimal model management for multifidelity Monte Carlo estimation},
  author    = {B. Peherstorfer and K. Willcox and M. Gunzburger},
  journal   = {SIAM Journal on Scientific Computing},
  volume    = {38},
  number    = {5},
  pages     = {A3163--A3194},
  year      = {2016},
  publisher = {SIAM}
}

@article{plummer2015cuts,
  author    = {Plummer, M.},
  journal   = {Statistics and Computing},
  number    = {1},
  pages     = {37--43},
  publisher = {Springer},
  title     = {{Cuts in Bayesian graphical models}},
  volume    = {25},
  year      = {2015}
}

@article{prime:2026,
  title     = {Vanishing dynamic strength measured in a transient high-pressure phase of tin},
  author    = {M. B. Prime and J. L. Brown and S. J. Fensin and D. R. Jones and J. W. Dyer},
  journal   = {Phys. Rev. E},
  volume    = {113},
  issue     = {6},
  pages     = {065502},
  year      = {2026},
  month     = {Jun},
  publisher = {American Physical Society},
  doi       = {10.1103/q5kp-5yj1}
}

@article{ramsay-li:1998,
  author  = {J. O. Ramsay and X. Li},
  journal = {Journal of the Royal Statistical Society, Ser. B},
  number  = {2},
  pages   = {351-363},
  title   = {Curve registration},
  volume  = {60},
  year    = {1998}
}

@book{ramsay2005,
  author    = {Ramsay, J. O. and Silverman, B. W.},
  publisher = {Springer},
  title     = {Functional Data Analysis},
  year      = {2005}
}

@inproceedings{Rob06a,
  author    = {Robinson, T.~D. and Eldred, M.~S. and Willcox, K.~E. and Haimes, R.},
  title     = {Strategies for Multifidelity Optimization with Variable Dimensional Hierarchical Models},
  booktitle = {Proceedings of the 47th AIAA/ASME/ASCE/AHS/ASC Structures, Structural Dynamics, and Materials Conference (2nd AIAA Multidisciplinary Design Optimization Specialist Conference)},
  year      = {2006},
  address   = {Newport, RI},
  month     = {May 1--4,},
  note      = {AIAA Paper 2006-1819}
}

@inproceedings{Rob06b,
  author    = {Robinson, T.~D. and Willcox, K.~E. and Eldred, M.~S. and Haimes, R.},
  title     = {Multifidelity Optimization for Variable-Complexity Design},
  booktitle = {Proceedings of the 11th AIAA/ISSMO Multidisciplinary Analysis and Optimization Conference},
  year      = {2006},
  address   = {Portsmouth,~VA},
  month     = {September~6--8,},
  note      = {AIAA Paper 2006-7114}
}

@article{rumsey2025all,
  title   = {All Emulators are Wrong, Many are Useful, and Some are More Useful Than Others: A Reproducible Comparison of Computer Model Surrogates},
  author  = {Rumsey, K. N. and Gibson, G. C. and Francom, D. and Morris, R.},
  journal = {arXiv preprint arXiv:2512.09060},
  year    = {2025}
}

@article{sacks1989SS,
  title     = {Design and analysis of computer experiments},
  author    = {Sacks, J. and Welch, W. J. and Mitchell, T. J. and Wynn, H. P.},
  journal   = {Statistical Science},
  pages     = {409--423},
  year      = {1989},
  publisher = {JSTOR}
}

@inproceedings{savage2007,
  author        = {Savage, M. E. and Bennett, L. F. and Bliss, D. E. and Clark, W. T. and Coats, R. S. and Elizondo, J. M. and LeChien, K. R. and Harjes, H. C. and Lehr, J. M. and Maenchen, J. E. and others},
  booktitle     = {2007 16th IEEE International Pulsed Power Conference},
  doi           = {10.1103/PhysRevSTAB.13.010402},
  organization  = {IEEE},
  pages         = {979--984},
  title         = {An overview of pulse compression and power flow in the upgraded {Z} pulsed power driver},
  volume        = {2},
  year          = {2007}
}

@article{siam_ml_mcmc,
  author  = {Dodwell, T. J. and Ketelsen, C. and Scheichl, R. and Teckentrup, A. L.},
  title   = {Multilevel Markov Chain Monte Carlo},
  journal = {SIAM Review},
  volume  = {61},
  number  = {3},
  pages   = {509-545},
  year    = {2019},
  doi     = {10.1137/19M126966X}
}

@article{srivastava-etal-JASA:2011,
  author  = {A. Srivastava and W. Wu and S. Kurtek and E. Klassen and J. S. Marron},
  journal = {arXiv:1103.3817v2 [math.ST]},
  title   = {Registration of Functional Data Using {F}isher-{R}ao Metric},
  year    = {2011}
}

@book{srivastava2016,
  author    = {A. Srivastava and E. Klassen},
  publisher = {Springer},
  title     = {Functional and Shape Data Analysis},
  year      = {2016}
}

@article{tin_llnl,
  author  = {Johnson, J. N. and Hixson, R. S. and Gray, G. T., III and Morris, C. E.},
  title   = {Quasielastic release in shock‐compressed solids},
  journal = {Journal of Applied Physics},
  volume  = {72},
  number  = {2},
  pages   = {429-441},
  year    = {1992},
  month   = {07},
  doi     = {10.1063/1.351871}
}

@article{tucker-wu-srivastava:2013,
  author  = {J. D. Tucker and W. Wu and A. Srivastava},
  journal = {Computational Statistics and Data Analysis},
  pages   = {50-66},
  title   = {Generative models for functional data using phase and amplitude separation},
  volume  = {61},
  year    = {2013}
}

@article{walters2018bayesian,
  author    = {Walters, D. J. and Biswas, A. and Lawrence, E. C. and Francom, D. C. and Luscher, D. J. and Fredenburg, D. A. and Moran, K. R. and Sweeney, C. M. and Sandberg, R. L. and Ahrens, J. P. and others},
  journal   = {Journal of Applied Physics},
  number    = {20},
  pages     = {205105},
  publisher = {AIP Publishing},
  title     = {Bayesian calibration of strength parameters using hydrocode simulations of symmetric impact shock experiments of {Al-5083}},
  volume    = {124},
  year      = {2018}
}

@article{williams2006combining,
  author    = {Williams, B. and Higdon, D. and Gattiker, J. and Moore, L. and McKay, M. and Keller-McNulty, S. and others},
  journal   = {Bayesian Analysis},
  number    = {4},
  pages     = {765--792},
  publisher = {International Society for Bayesian Analysis},
  title     = {Combining experimental data and computer simulations, with an application to flyer plate experiments},
  volume    = {1},
  year      = {2006}
}

\end{CJK}

\end{document}